\pdfoutput=1

\documentclass[apj,twocolumn]{openjournal}
\usepackage{amsmath}
\usepackage{gensymb}
\usepackage{booktabs}
\usepackage{soul}

\usepackage[dvipsnames]{xcolor} 
\usepackage[breaklinks,colorlinks,citecolor=blue,urlcolor=blue,linkcolor=blue]{hyperref}

\usepackage{orcidlink}

\renewcommand{\arraystretch}{1.2}  
\usepackage{listings}
\usepackage{color}

\usepackage{soul}
\usepackage{amsmath}
\usepackage{xspace}
\usepackage{graphicx}
\usepackage{enumitem}
\usepackage{amssymb}
\usepackage{xifthen}
\usepackage{hyperref}
\usepackage[normalem]{ulem}
\usepackage{multirow}

\usepackage[hang,flushmargin]{footmisc} 

\makeatletter
\newcommand{\unit}[1]{%
    \,\mathrm{#1}\checknextarg}
\newcommand{\checknextarg}{\@ifnextchar\bgroup{\gobblenextarg}{}}
\newcommand{\gobblenextarg}[1]{\,\mathrm{#1}\@ifnextchar\bgroup{\gobblenextarg}{}}
\makeatother

\newif\ifstartedinmathmode
\newcommand{\msun}{%
  \relax\ifmmode\startedinmathmodetrue\else\startedinmathmodefalse\fi
  {\ifstartedinmathmode\unit{M_{\odot}}\else$\unit{M_{\odot}}$\fi}\xspace%
}

\newif\ifstartedinmathmode
\newcommand{\rsun}{%
  \relax\ifmmode\startedinmathmodetrue\else\startedinmathmodefalse\fi
  {\ifstartedinmathmode\unit{R_{\odot}}\else$\unit{R_{\odot}}$\fi}\xspace%
}

\makeatletter
\renewcommand\@makecaption[2]{%
  \par
  \vskip\abovecaptionskip
  \begingroup
    \footnotesize\rmfamily
    \begingroup
      \samepage
      \flushing
      \let\footnote\@footnotemark@gobble
      \ifnum\pdfstrcmp{\@captype}{table}=0
        \@make@capt@title{\textsc{Table \thetable}}{#2}%
      \else
        \ifnum\pdfstrcmp{\@captype}{figure}=0
          \@make@capt@title{\textsc{Figure \thefigure}}{#2}%
        \else
          \@make@capt@title{#1}{#2}%
        \fi
      \fi\par
    \endgroup
  \endgroup
  \vskip\belowcaptionskip
}
\makeatother

\begin{document}

\author{Ryan Eskenasy\,\orcidlink{0000-0003-0854-8313}$^{1}$}
\author{Valeria Olivares\,\orcidlink{0000-0001-6638-4324}$^{2,3}$}
\author{Yuanyuan Su\,\orcidlink{0000-0002-3886-1258}$^{1}$}

\affiliation{$^1$Department of Physics and Astronomy, University of Kentucky, 505 Rose Street, Lexington, KY 40506, USA}

\affiliation{$^2$Departamento de F\'isica, Universidad de Santiago de Chile, Av. Victor Jara 3659, Santiago 9170124, Chile}

\affiliation{$^3$Center for Interdisciplinary Research in Astrophysics and Space Exploration (CIRAS), Universidad de Santiago de Chile, Santiago 9170124, Chile}

\email{Corresponding author: reskenasy@uky.edu}
\title{The Rise of Ionized Gas Filaments in Early-Type Galaxies}

\bigskip
\begin{abstract}
    Multiphase filamentary nebulae are ubiquitous in the brightest cluster galaxies (BCGs) of cool-core clusters, providing insight into baryon cycling and the feeding and feedback of supermassive black holes. However, BCGs account for less than 1\% of all early-type galaxies (ETGs). To broaden our understanding of how multiphase filamentary nebulae form in ETGs and connect to the greater picture of galaxy evolution, it is crucial to explore ETGs that are outside of the dense centers of galaxy clusters or groups. We present VLT-MUSE IFU observations of 126 nearby non-central ETGs, detecting warm ionized gas in 62 of them. 35/62 host rotating gas disks with the majority of them morphologically and kinematically aligned with their stellar components, suggesting stellar mass loss may dominate their warm-gas origin. The remaining 27 host filamentary nebulae, often decoupled from the stellar components, resembling those observed in BCGs. These filamentary sources display unique emission line properties that cannot be fully explained by photoionization from post-asymptotic giant branch stars, active galactic nuclei, or fast gas shocks alone. For the twelve filamentary sources that have \emph{Chandra} data, their soft X-ray emission exceeds or is consistent with (within uncertainties) unresolved low-mass X-ray binary emission, indicating that filamentary systems generally host an appreciable hot gas reservoir. We suggest that cooling-related processes, e.g., self-irradiation associated with the cooling hot gas, may contribute to the powering of warm gas line emission, similar to cool-core clusters, though the detailed mechanisms and physical conditions may differ. As a case study, we investigate NGC~4374, a non-central ETG with extensive \emph{Chandra} observations, and find that its warm filaments are over-pressured compared to the hot filaments -- opposite to what is observed in cool-core clusters. 
\end{abstract}

\maketitle
\section{Introduction}

Early-type galaxies (ETGs), comprising elliptical and lenticular galaxies, were long believed to be devoid of gas species that are necessary for star-formation (SF). Their characteristically low star-formation rates (SFRs) led to the prevailing view that ETGs were passive, ``red and dead'' systems \citep{bregman1978galactic,mathews1971galactic}. However, such understanding of ETGs has been recently challenged due to the advent of telescopes with superior sensitivity and resolution. Gas that is multiphase in temperature and density, including (but not limited to) cold molecular ($T<100$ K) \citep{edge2001detection,salome2003cold,young2011atlas3d,tremblay2018galaxy,olivares2019ubiquitous}, warm ionized ($T\sim10^{4-5}$ K) \citep{heckman1989dynamical,hatch2006origin,hamer2016optical,olivares2019ubiquitous}, and diffuse hot ($T\sim10^{7-8}$ K) gas \citep{RevModPhys.58.1,fabbiano1989x,su2015scatter} have been detected. The exciting detections of multiphase gas well motivate the astronomical community to revisit models describing the mass-assembly history of ETGs and their morphological evolution, given their minimal SFR.

Observations of massive galaxies sitting in the center of galaxy clusters with sharply peaked X-ray surface brightness profiles (hereafter cool-core clusters) provide valuable insight into this puzzle. Their diffuse, hot intracluster medium (ICM) loses energy primarily through thermal bremsstrahlung as a result of free charges interacting with ions. For a spherical hot halo in quasi-hydrostatic equilibrium, there is a radius at which the cooling time (that is, the timescale for gas to radiate away its thermal energy content) is roughly the age of the system. Within this radius, radiative cooling drives an increase in gas density to maintain pressure support of the above gas layers. The increase in density is only plausible if there is an inward flow of gas funneling towards the center of the gravitational potential well, where typically the brightest cluster galaxy (BCG) resides. However, expected signatures such as X-ray coolant lines and high SFR are not observed. Thus, we are left with the ``cooling-flow'' problem which lies at the forefront of unresolved astrophysical puzzles \citep{lea1973thermal,fabian1977subsonic,mathews1978radiative,nulsen1986thermal,fabian1994cooling}.

A heating source has been proposed to explain the lack of SF in these systems. An enticing candidate is the BCG's active galactic nucleus (AGN), a highly energetic region driven by gas accretion onto the supermassive black hole (SMBH), which has a typical mass of $\sim10^{6-10}~\mathrm{M}_{\odot}$. In this scenario, some of the accreting rest-mass energy is released in radiation and kinetic outflows, and is enough to counter SF. Various lines of evidence suggest that cool-core clusters exhibit this phenomenon. For example, X-ray observations of the hot gas around BCGs in clusters with low core entropy ($\lesssim$ 30 keV cm$^2$) and short central cooling times ($\lesssim$ 1 Gyr) reveal regions of depressed surface brightness \citep{mcnamara2000chandra,fabian2000chandra,wise2007x,cavagnolo2008entropy,dunn2008investigating}. Radio observations find that these areas are also often filled with relativistic radio plasma \citep{mcnamara2005heating,vantyghem2014cycling}. It is expected that AGN jets excavate these pockets of diffuse hot gas, and impart energy on their surroundings, thus preventing cooling and inhibiting SF (though, the exact mechanism that thermalizes the atmosphere is unknown). Aside from these clues provided by X-ray and radio observations, sub-mm and optical spectra show extended nebular emission in cool-core clusters \citep{edge2001detection,salome2003cold,cavagnolo2008entropy,rafferty2008regulation,olivares2019ubiquitous}. Such findings have emphasized the role that multiphase emission line nebulae may play in explaining the connection between AGN feedback and SF.

Atacama Large Millimeter Array (ALMA) and VLT Multi Unit Spectroscopic Explorer (MUSE) observations have revolutionized our understanding of the multiphase gas properties in massive ETGs. ALMA and MUSE provide a thorough view of the cold molecular gas (often traced by CO rotational lines) and warm ionized gas (typically traced by the H$\alpha$ recombination line), respectively. The exceptional spatial and spectral resolution of these facilities allow their morphologies and kinematics to be viewed in great detail. In fact, ALMA observations of cool-core clusters find cold molecular gas in the form of filamentary networks, that are co-spatial and co-moving with the warm ionized gas filaments detected by MUSE \citep{tremblay2018galaxy,olivares2019ubiquitous,olivares2022gas}. This coupling between gas phases strongly suggests that they are linked. Condensation models suggest that the warm filaments envelope the cold molecular filaments, as these phases ultimately condense out of the thermally unstable hot halo. These multiphase nebulae often wrap around the aforementioned X-ray cavities, which could be explained by the bubbles buoyantly rising, and in doing so adiabatically lifting the low-entropy gas to radii where thermal instabilities are induced \citep{revaz2008formation,li2014modeling}.

Studies of the most massive ETGs—such as brightest cluster galaxies (BCGs) and brightest group galaxies (BGGs; e.g., \citealp{olivares2022gas})—provide important but narrow insights into the formation of multiphase gas. BCGs represent only a small subset of the ETG population and occupy unique, extreme environments. They typically reside in the hot, dense cores of their host clusters’ intracluster medium (ICM). In these locations, the high ambient gas density makes it difficult to isolate and probe the truly low-entropy gas associated with cooling. Moreover, BCGs likely have complex evolutionary histories, often shaped by multiple mergers with surrounding satellites, further complicating the interpretation of observations. Extending spatially resolved studies, using integral-field
unit (IFU) spectroscopic observations, to systems located away from the extreme centers of clusters and groups (hereafter “non-central” ETGs) will broaden our understanding of multiphase gas formation in ETGs and contribute to the larger picture of galaxy evolution.

Indeed large IFU surveys such as ATLAS$^{\mathrm{3D}}$ \citep{cappellari2011atlas3d}, CALIFA \citep{sanchez2012califa,walcher2014califa}, SAMI \citep{croom2012sydney,bryant2015sami}, and MaNGA \citep{bundy2014overview} have explored the spatially resolved properties of nearby galaxies, including non-central ETGs. These surveys have transformed our knowledge of key properties that were largely inaccessible with single-slit spectroscopy, including but not limited to metallicity gradients, baryonic and dark matter distributions, and region-dependent ionization mechanisms. Despite this significant progress, these large surveys are not optimized to probe the detailed multiphase gas structure. In particular, their spatial resolution and wavelength coverage are generally insufficient to fully resolve filamentary warm gas structures and to exploit a complete set of optical emission-line-diagnostics. For example, while the SAURON spectrograph, used in the ATLAS$^{\mathrm{3D}}$ survey, has detected warm ionized gas in NGC~4374 \citep{sarzi2006sauron}, its narrow field of view and small wavelength coverage (which excludes rest-frame H$\alpha$) preclude a truly detailed mapping of the nebulae. Spatially and spectrally resolving such structures is a necessary step toward understanding the role of cooling in producing filamentary multiphase gas formation in non-central ETGs, and can be accomplished with a facility like MUSE.

\cite{eskenasy2024formation} analyzed the multiphase gas properties of 16 nearby, non-central ETGs selected from the ATLAS$^{\mathrm{3D}}$ survey using archival MUSE and \emph{Chandra} X-ray observations. In this sample, the majority of sources with H$\alpha$ emission have warm gas in the form of rotating disks, unlike the massive BCGs. Many of these rotating gas disks are kinematically coupled with the stars, suggesting that a significant amount of their warm gas may have formed from the shedding of their older stellar population. Only three systems in that sample show filamentary nebulae, and each of these three do not show coupling between their stars and gas. Interestingly, the \emph{Chandra} observations reveal that the only sources with signatures of hot gas reservoirs were those with filamentary warm gas. This is consistent with the prevailing model in which the observed H$\alpha$ filaments condense out of a thermally unstable hot phase, analogous to BCGs, where warm gas filaments are generally interpreted as having cooled from the hot ICM. Although this study, as well as others \citep{temi2022probing}, have paved the way for assessing multiphase gas formation and AGN feedback in non-central ETGs, the sample size is limited. Only two of the three ETGs in \cite{eskenasy2024formation} with H$\alpha$ filaments resemble the extended filamentary networks observed in BCGs. The literature lacks an adequate sample size of filamentary emission-line nebulae, to properly test the feasibility of cooling in typical ETGs.

In this study, we aim to analyze the multiphase gas properties of 126 nearby, non-central ETGs. Archival MUSE and (when available) \emph{Chandra} observations will assess their warm ionized and hot gas content, respectively. This statistically significant sample of non-central ETGs will serve to be a stepping-stone in filling in the literature gap when it comes to understanding the origin of multiphase, emission-line nebulae in systems other than the most massive, cluster and group central galaxies. Such a study is required to form a complete understanding of baryon cycling and AGN feedback in ETGs across the entire mass scale.

In Sec.~\ref{sec:sample} we describe the parent sample from which we adopt our non-central ETGs from. Sec.~\ref{sec:data} describes the procedures and tools use to reduce the MUSE and \emph{Chandra} data. These data are described and discussed in Sec.~\ref{sec:results} and Sec.~\ref{sec:discuss}, respectively. Concluding remarks and a full summary are provided in Sec.~\ref{sec:conclusion}.

\bigskip
\section{Sample Description}
\label{sec:sample}

\begin{figure}
    \centering
    \includegraphics[width=0.99\linewidth]{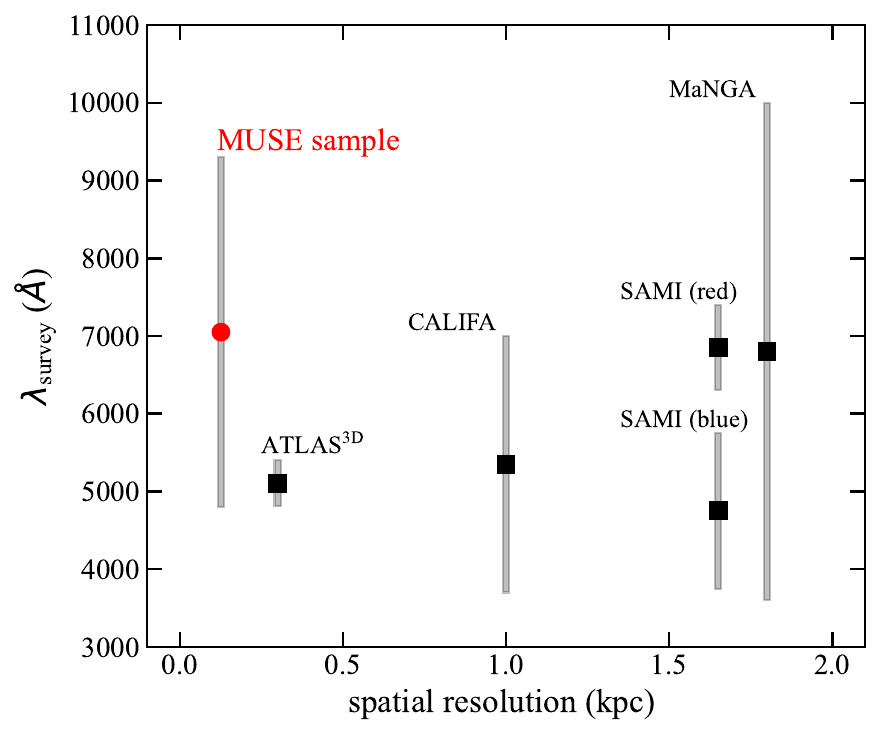}
    \caption{Distribution of spectral coverage and median physical spatial resolution of our MUSE sample (red circle) compared to the ATLAS$^{\mathrm{3D}}$, CALIFA, SAMI, and MaNGA IFU surveys (black squares). The sub-kpc resolution and broad optical wavelength coverage of our data allow for a detailed view of the warm ionized phase of the ISMs of our sample galaxies.}
    \label{fig:IFU}
\end{figure}

In this paper we study 126 nearby ($D\lesssim46$ Mpc) ETGs, a subset of the 50 Mpc Galaxy Catalog (50MGC; \citealp{ohlson202350}). 50MGC consists of 15,424 galaxies within 50 Mpc and is the first uniform and homogeneous source of general galaxy properties at this distance. The catalog's self-consistent measurements (including morphological classification, mass, distance, etc.) allow for a broad range of analyses of local galaxies. The close
proximity to these galaxies permits detailed observations of their stellar and gaseous components.  Furthermore, the spatial resolution of MUSE allows us to explore the nebulae within our sample on sub-kpc scales. Fig.~\ref{fig:IFU} compares the spatial resolution and wavelength coverage of our sample with that of the ATLAS$^{\mathrm{3D}}$, CALIFA, SAMI, and MaNGA IFU surveys - which have transformed our understanding of the gaseous properties of galaxies. Our sample can build on this progress by combining enhanced spatial resolution with an average spectral resolution $\sim3000$, sufficient for identifying key, individual diagnostic emission lines.

To extract the aforementioned sub sample of 126 ETGs, we performed the following filters:

\begin{enumerate}
\item Sources classified as ETGs according to 50MGC's morphological reclassification scheme. ETGs are identified as having \verb|best_dist| = ``early''.
\item Sources with stellar mass estimates indicated by the \verb|logmass| column. These stellar masses were derived by translating the $g-i$ vs. $M/L_i$ relation from \cite{taylor2011galaxy} to the available colors within the catalog.
\item Sources with archival MUSE observations that have pointings within 1$^{\prime}$ of the catalog's published coordinates (\verb|ra| and \verb|dec| columns). We remove sources captured in Narrow-Field Mode since its small field of view (7.42$^{\prime\prime}~\times~$7.42$^{\prime\prime}$) prevents full spatial coverage of the galaxies at this distance.
\end{enumerate}

In addition to these cuts made using the catalog's published data, we manually removed 32 ETGs that were classified as either BGGs or BCGs in the literature. A large portion of removed BGGs are those within the Complete Local-volume Group Sample \citep{o2017complete}. Performing these cuts results in our sub sample comprised of 126 ETGs sitting outside of dense group and cluster cores, as well as some in truly isolated environments. This statistically significant sample of nearby, non-central ETGs is ideal for testing the internal physics driving multiphase gas formation across a broad stellar mass range (6.20~$\lesssim~$$\text{log}(\text{M}_{\star}/\text{M}_{\odot})$$~\lesssim11.3$). The top panel of Fig.~\ref{fig:hist} shows the stellar mass distribution of all ETGs as well as late-type galaxies (LTG) within the 50MGC sample (with available stellar mass estimates). The overlaid solid black histogram shows the stellar mass distribution of our subsample described above.

\bigskip
\section{Observations \& Data Reduction} 
\label{sec:data}

\subsection{MUSE Optical Integral Field Spectroscopy} 
\label{subsec:muse}

Archival MUSE IFU observations were used to deduce the nebular and stellar properties of our sample of ETGs. Table.~\ref{tab:muse} provides the observation log for all sources in our sample. MUSE is a wide field of view (1$^{\prime}~\times~$1$^{\prime}$), image-slicing IFU spectrograph capable of achieving exquisite spatial resolution. The data were reduced using the European Southern Observatory Recipe Execution Tool (ESOREX v.3.13.1; \citealp{weilbacher2020data}). The final datacube was sky subtracted using version 2.1 of the Zurich Atmosphere Purge (ZAP; \citealp{soto2016zap}) in addition to the data reduction pipeline's internal sky subtraction method. The data are corrected for galactic extinction using the dust maps of \cite{schlegel1998maps}. The final MUSE datacube covers wavelengths between 4750-9300 \text{\AA}.

Higher level data products were created by fitting the stellar continuum and nebular emission lines. Using template matching and cross-correlation, \verb|AUTOZ| \citep{baldry2014galaxy} estimated the redshift of the star light in each spaxel. The \verb|VDISPFIT| code then found the best-fit stellar velocity dispersions. The \verb|PLATEFIT| spectral fitting code \citep{brinchmann2004physical,tremonti2004origin} then independently fit the stellar continuum and nebular emission lines. The former was created by first masking possible nebular emission lines along the spectrum. Then, a series of stellar population templates were used to fit the underlying stellar continuum. This fitting procedure uses the stellar redshift and velocity dispersion maps mentioned above. Subtracting off the fitted stellar continuum results in a residual spectrum that is used to fit the nebular emission lines. All emission lines were fitted assuming a single Gaussian profile, and were tied to have similar velocities and velocity dispersions. 

To remove noise from the emission line maps, we exclude pixels with a signal-to-noise ratio $<7$ and gas velocity dispersions $<40$ km s$^{-1}$. Further pixels are removed if their 1$\sigma$ velocity and velocity dispersion uncertainties exceed a given threshold. This cut-off was determined on a source-by-source basis by calculating the average velocity dispersion uncertainty in an offset region.

\medskip
\subsubsection{Stellar kinematics}
\label{subsec:gist}

Using the Galaxy IFU Spectroscopy Tool pipeline (GIST; \citealp{bittner2019gist}), the stellar kinematics were extracted from the MUSE datacube. Spaxels were tessellated with the Voronoi SNR method outlined in \cite{cappellari2003adaptive}, to a target SNR$~=~$40 per bin. For galaxies with low surface brightness, this target could not be met and thus was lowered accordingly down to SNR$~\ge~$15. Spectra
were then logarithmically re-binned in wavelength. The \verb|pPXF| routine \citep{cappellari2004parametric} produced line-of-sight velocity and velocity dispersion maps for an initial velocity dispersion guess and rest-frame wavelength range. This wavelength range was determined on a source-by-source basis by searching for a region in the spectrum void of emission or absorption lines. Foreground stars were manually excluded using a spatial mask on the final maps.

\subsection{\it{Chandra} X-ray Data}
\label{subsec:chandrared}

We utilize \emph{Chandra} X-ray observations of 74 out of 126 ETGs in our sample. The \emph{Chandra} observation log for these ETGs is provided in Table.~\ref{tab:chandra}. This data was reduced using CIAO version 4.16 \citep{fruscione2006ciao} and CALDB version 4.10.2. First, the \verb|chandra_repro| routine reprocessed the Level 1 data. \verb|acisreadcorr| then corrected for out-of-time events from X-ray bright sources. Anomalous ACIS chip background flares were removed using \verb|deflare|. Data from separate observations were reprojected to a matching coordinate using the \verb|reproject_obs| routine, followed by merging of the event files. \verb|Blanksky| files were used for background sky subtraction. Exposure maps were created using the \verb|flux_obs| routine. Since we are interested in interpreting the diffuse gas properties, we removed point sources which were identified using \verb|wavdetect|. Regions with removed point sources were replaced with a region adjacent to the point source so that it contains the ISM emission and all sources of background. The count map (subtracted by the background file) was then normalized by the exposure maps resulting in our analysis-ready 0.5-2.0 keV image in units of counts per second.
\begin{figure}
    \centering
    \includegraphics[width=0.99\linewidth]{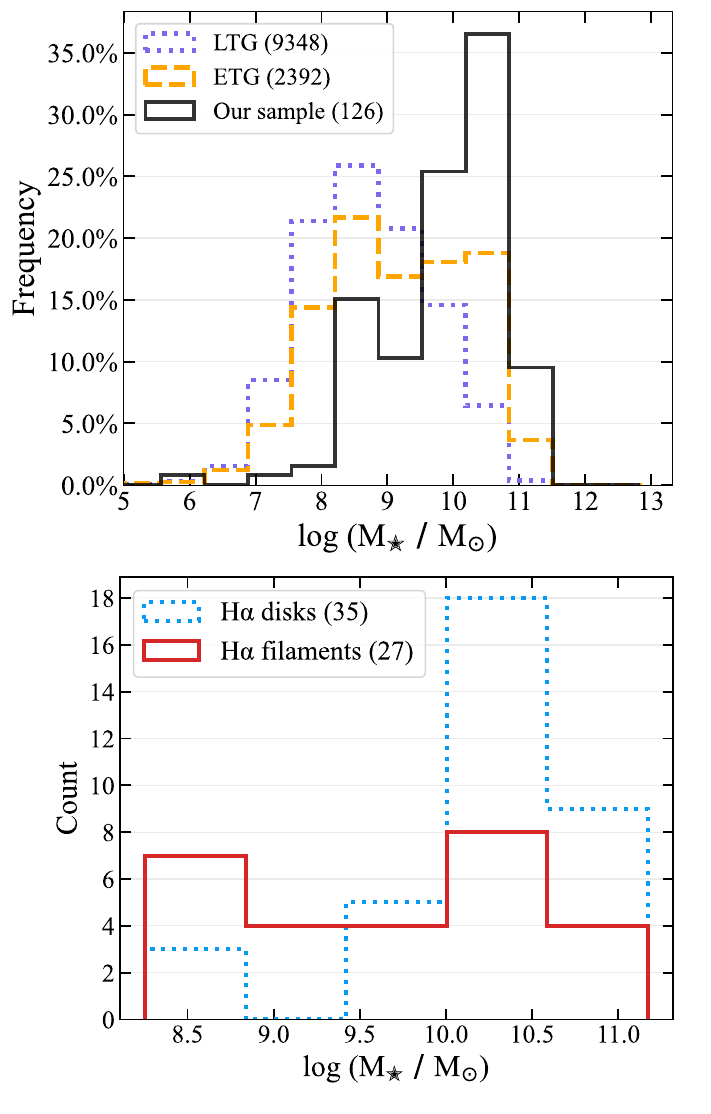}
    \caption{Top: Frequency histogram showing the full stellar mass distributions of early (orange dashed) and late (purple dotted)-type galaxies within the 50MGC catalog. The solid black histogram shows the stellar mass distribution of our subsample described in Sec.~\ref{sec:sample}. Bottom: Total count histogram for the sources within our sample with H$\alpha$ emission detected in their MUSE spectra. The red solid (blue dotted) histogram depicts sources with detected H$\alpha$ in the form of filaments (rotating disks). The total number of galaxies belonging to each group is provided in the legend of both panels.}
    \label{fig:hist}
\end{figure}
\subsubsection{X-ray luminosities}
\label{subsec:chandraanalysis}

To extract the 0.5-2.0 keV luminosities for each source, we extracted counts within a circular region centered on the source with a radius of one effective radius (R$_e$). We derived R$_e$ using the stellar mass-size relation given in Eq. 1 in \cite{fernandez2013stellar}. To convert our net count rates to energy fluxes, we used the PIMMS Mission Count Rate Simulator\footnote{\url{https://cxc.harvard.edu/toolkit/pimms.jsp}}. We assumed an absorbed plasma APEC model, {\tt phabs}$\times${\tt apec}, with solar metallicity and $kT=0.5$ keV\footnote{We assume $kT=0.87$ keV and $kT=0.76$ keV for NGC~1266 and NGC~4374, respectively \citep{su2015scatter}}. These priors are typical for the hot ISM of ETGs \citep{su2013investigating}. We refer to \cite{eskenasy2024formation} for describing how varying the metallicity and temperature impacts the X-ray luminosity using this spectral model. HI column densities were gathered from the HEASARC \href{https://heasarc.gsfc.nasa.gov/cgi-bin/Tools/w3nh/w3nh.pl}{$N_H$ tool} \citep{bekhti2016hi4pi} to account for absorption along the line of sight.

\section{Results} 
\label{sec:results}

The MUSE data reveals 62/126 ($\sim49\%$) non-central ETGs with H$\alpha$ emission in their spectrum, tracing warm ionized gas. These nebulae are divided into two general categories, based on their morphology and kinematics: rotating disks and filamentary networks. The histogram presented in the bottom panel of Fig.~\ref{fig:hist} shows the count and stellar mass distribution of these two morphological classifications. The full stellar mass range of all sources with H$\alpha$ emission span 8.25 $\lesssim$ log(M$_{\star}$/M$_{\odot}$) $\lesssim$ 11.17. The warm gas morphology and stellar mass of all 62 non-central ETGs are provided in Table.~\ref{tab:results}.

\subsection{Rotating warm gas disks}
\label{subsec:disk_analysis}

The rotating gas disks dominate the H$\alpha$ morphological distribution in our sample. We find that 35 out of 62 H$\alpha$-emitters exhibit rotating disks, which we define as warm nebulae showing a clear velocity gradient indicative of ordered rotation. The sizes of the gas disks, defined as the projected length of the disk's semi-major axis, range from roughly 0.1 - 8 kpc. 
\begin{figure*}

    \includegraphics[width=0.93\linewidth]{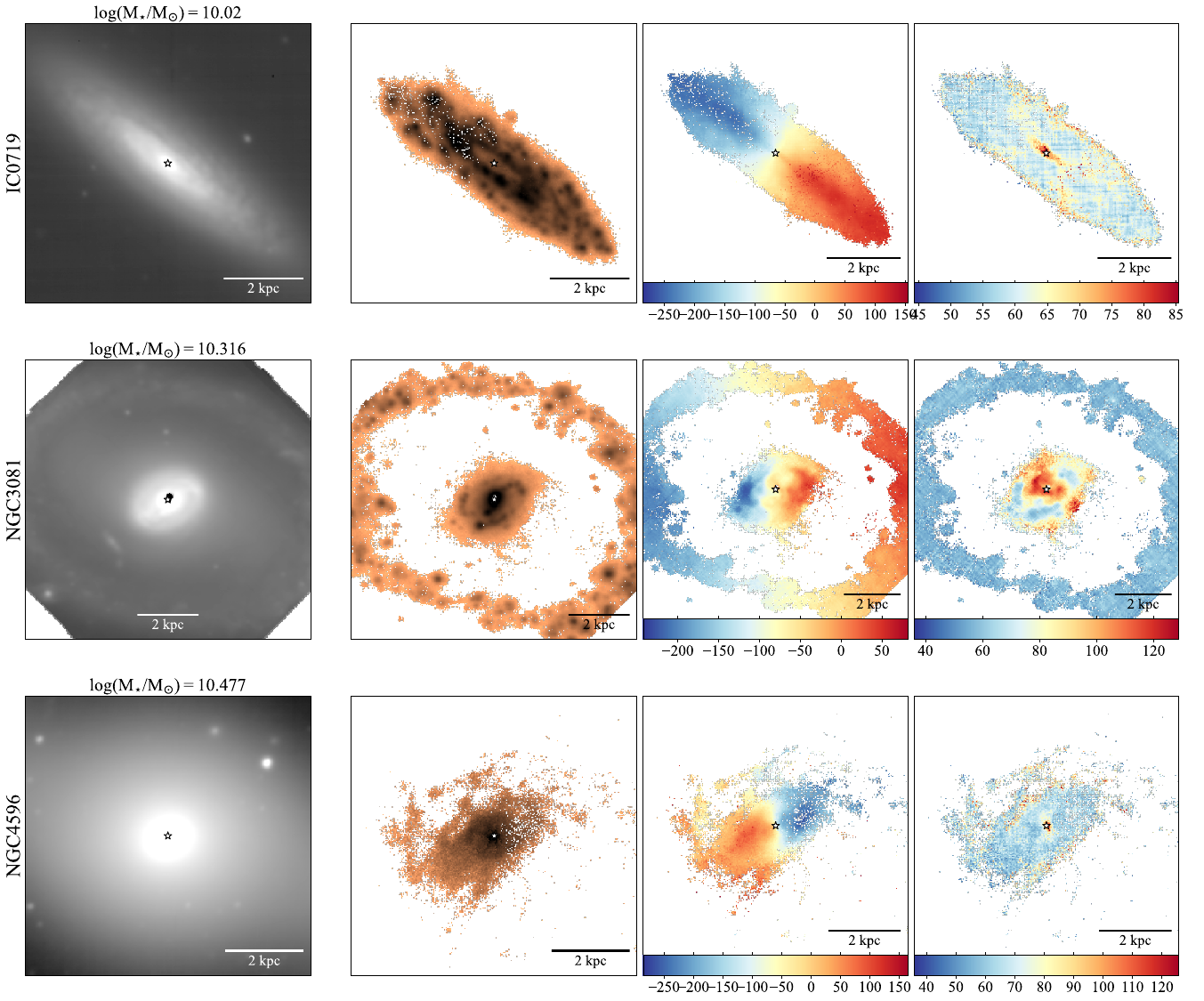}
    \caption{Three example ETGs with rotating H$\alpha$ disks. Each row corresponds to one of the three general subclasses of rotating disks we find in our sample: symmetric, volume filling disks (top), smaller nuclear disk surrounded by an outer ring (middle), and asymmetric disks with features likely indicating interaction and tidal effects (bottom). The name (stellar mass) of the galaxy is given along the side (top) of the leftmost panel. The first, second, third, and fourth columns correspond to the optical continuum, H$\alpha$ flux, warm gas LOS velocity (km s$^{-1}$), and warm gas velocity dispersion (km s$^{-1}$), respectively. The physical scale is given in the bottom right of each panel. The star denotes the catalog's published RA and Dec values.}
    \label{fig:disks}
\end{figure*}
Among these 35 ETGs, the disky morphologies are composed of three general types: (1) symmetric, volume-filling disks ($\sim34\%$). (2) smaller-scale nuclear disk, accompanied by a distinct outer ring constituted by several discrete clouds ($\sim17\%$). (3) asymmetric disks with irregular spatial distributions, perhaps driven by tidal effects ($\sim49\%$). Fig.~\ref{fig:disks} shows the optical continuum, H$\alpha$ flux, and the warm ionized gas line-of-sight velocity and velocity dispersion maps for three example ETGs in our sample that fall into the above categories.

\subsection{Warm gas filaments}
\label{subsec:fil_analysis}

The remaining 27 non-central ETGs with H$\alpha$ emission have warm gas morphologies in the form of filamentary networks. A source is classified as filamentary if its gas nebulae show compact or extended, irregular structures--typically extending from the optical center of the galaxy. Additionally their gas velocity fields are generally absent of clear, ordered rotation. 

The optical continuum, H$\alpha$ flux, velocity, and velocity dispersion for all 27 filamentary sources in our sample are shown in Fig.~\ref{fig:fils}. To quantify the size of the filaments, we refer to the projected distance from the optical center of the galaxy to its most radially extended filament. The size of the filaments range from 0.3 - 4 kpc.

\subsection{Stellar kinematics}
\label{subsec:stellar_kin}

Using the \emph{GIST} pipeline (see Sec.~\ref{subsec:gist}) we derive the line-of-sight velocity and velocity dispersion maps tracing the stellar population. $\sim79\%$ of our ETGs with H$\alpha$ emission show regular, ordered rotation in their stellar velocity fields. Low-mass (log(M$_{\star}$/M$_{\odot}$) $\lesssim$ 9.5) sources occupy $\sim69\%$ of the ETGs that are absent of ordered stellar rotation. For some of these low-mass ETGs, this may be attributed to their low stellar-surface brightness preventing the pipeline from adequately deriving the galaxy's underlying line-of-sight velocity distribution.

The H$\alpha$ and stellar velocity fields of ETGs hosting rotating H$\alpha$ disks in general show kinematic and spatial overlap between their gaseous and stellar disks. This is not the case for the H$\alpha$ filamentary sources. The asymmetric, non-rotating filamentary nebulae in many sources appear to be dissociated with their stellar population (see Sec.~\ref{subsec:stellar} and \ref{pagb} for further discussion).

\subsection{Hot gas content}
\label{subsec:hotgas}

High spatial-resolution X-ray observations are essential to probing the hot gas content of ETGs. ETGs are known to harbor low mass X-ray binaries (LMXBs) manifested as point sources, which can severely contaminate the study of their diffuse hot ISM. The superb spatial resolution of \emph{Chandra} ($\sim0.5^{\prime\prime}$) is well-suited to identify and eliminate a majority of point sources. 

\begin{figure}
    \centering
    \includegraphics[width=0.95\linewidth]{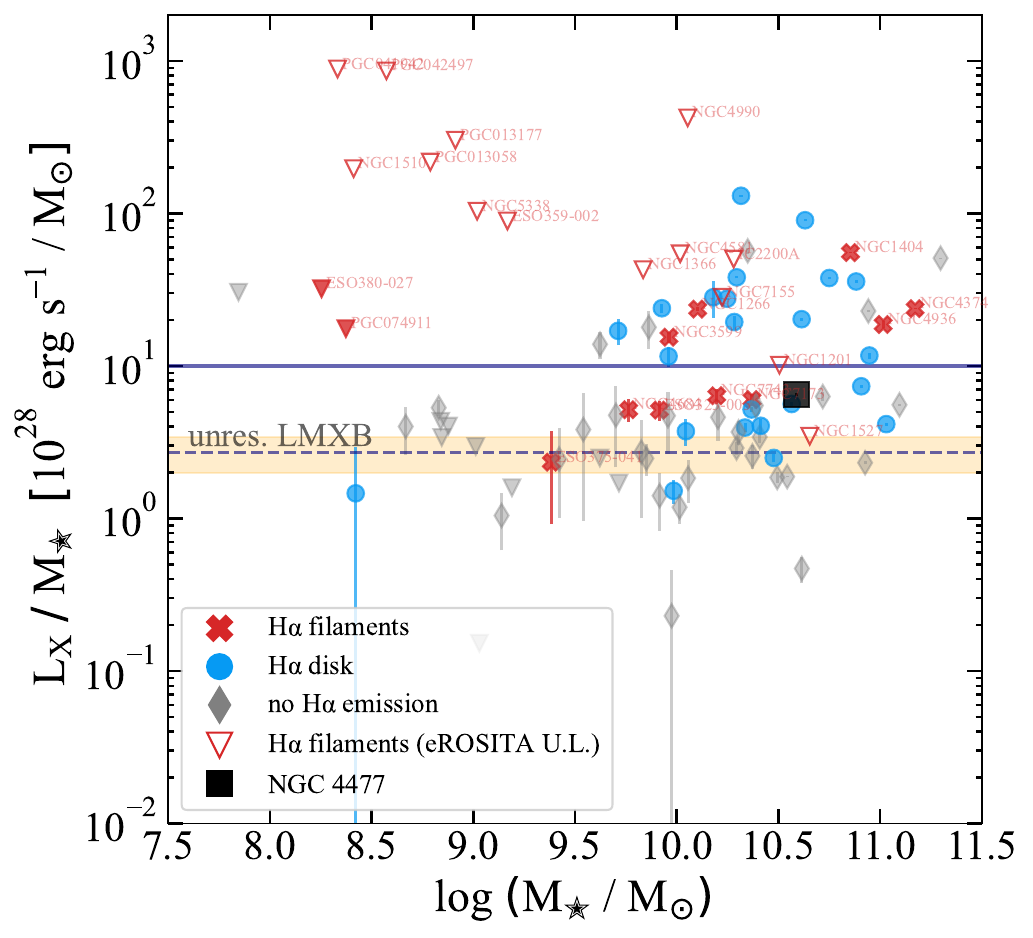}
    \caption{0.5-2.0 keV luminosity normalized by the stellar mass (vertical axis) vs. stellar mass (horizontal axis). Sources with filamentary H$\alpha$ nebulae are shown as red X's while rotating H$\alpha$ disks are blue circles. Sources without H$\alpha$ emission are depicted as gray diamonds. For very faint sources, \emph{Chandra} upper-limits are shown as filled inverted triangles. Open inverted triangles represent upper-limits on $L_X$ provided in eRASS1 for filamentary sources not observed with \emph{Chandra}. The black square represents NGC~4477 - the smallest known ETG with X-ray cavities and cooling signatures \citep{li2018x}. The solid blue line is a rough expectation of the emission from unresolved point sources likely influencing the eROSITA upper-limits. The dashed blue line denotes the estimated 0.5-2.0 keV emission from unresolved LMXBs as given in \citet{hou2021x}. We expect the ETGs sitting above this line to host diffuse, hot gas, and are thus systems where cooling is a feasible origin to their warm gas.}
    \label{fig:Lx}
\end{figure}

In Fig.~\ref{fig:Lx}, we compare the diffuse X-ray luminosity in 0.5-2.0\,keV (normalized by stellar mass) against the stellar mass for all ETGs in our sample with available MUSE and \emph{Chandra} data. The horizontal dashed line provides the average X-ray emissivity of unresolved LMXBs from \citet{hou2021x}. Galaxies sitting above this line have soft diffuse X-ray emission exceeding what is expected from unresolved LMXBs and likely contain a hot gas reservoir. We distinguish our data by those with filamentary ionized gas (red x's), ionized gas in the form of rotating disks (blue circles), and those where H$\alpha$ went undetected (gray diamonds). The inverted filled triangles show filamentary systems where the emissivity was computing using an upper-limit on $L_X$. Inverted open triangles show filamentary systems not observed by \emph{Chandra}, and their upper-limit on $L_X$ from eROSITA observations is gathered from eRASS1 \citep{predehl2021erosita,tubin2024erosita}\footnote{We homogenized the eROSITA catalog upper limits (originally reported in the 0.6-2.3 keV band, assuming an absorbed power-law spectrum with $\Gamma=2$ and $N_\mathrm{H}=3\times10^{20}$ cm$^{-2}$) by computing a bandpass/model conversion factor and then re-expressing them as 0.5-2.0 keV upper limits under our fiducial thermal model.}, when available. All ETGs with filamentary ionized gas are likely to contain hot gas, as are a strong percentage of the ETGs containing rotating gas disks. Roughly half of the ETGs where H$\alpha$ went undetected in emission contain hot gas. Therefore, we observe a mild connection between the hot gas content and the morphology of the warm gas, when compared at nearly the same mass range.

\begin{figure*}
  \centering
  \includegraphics[width=0.7\linewidth]{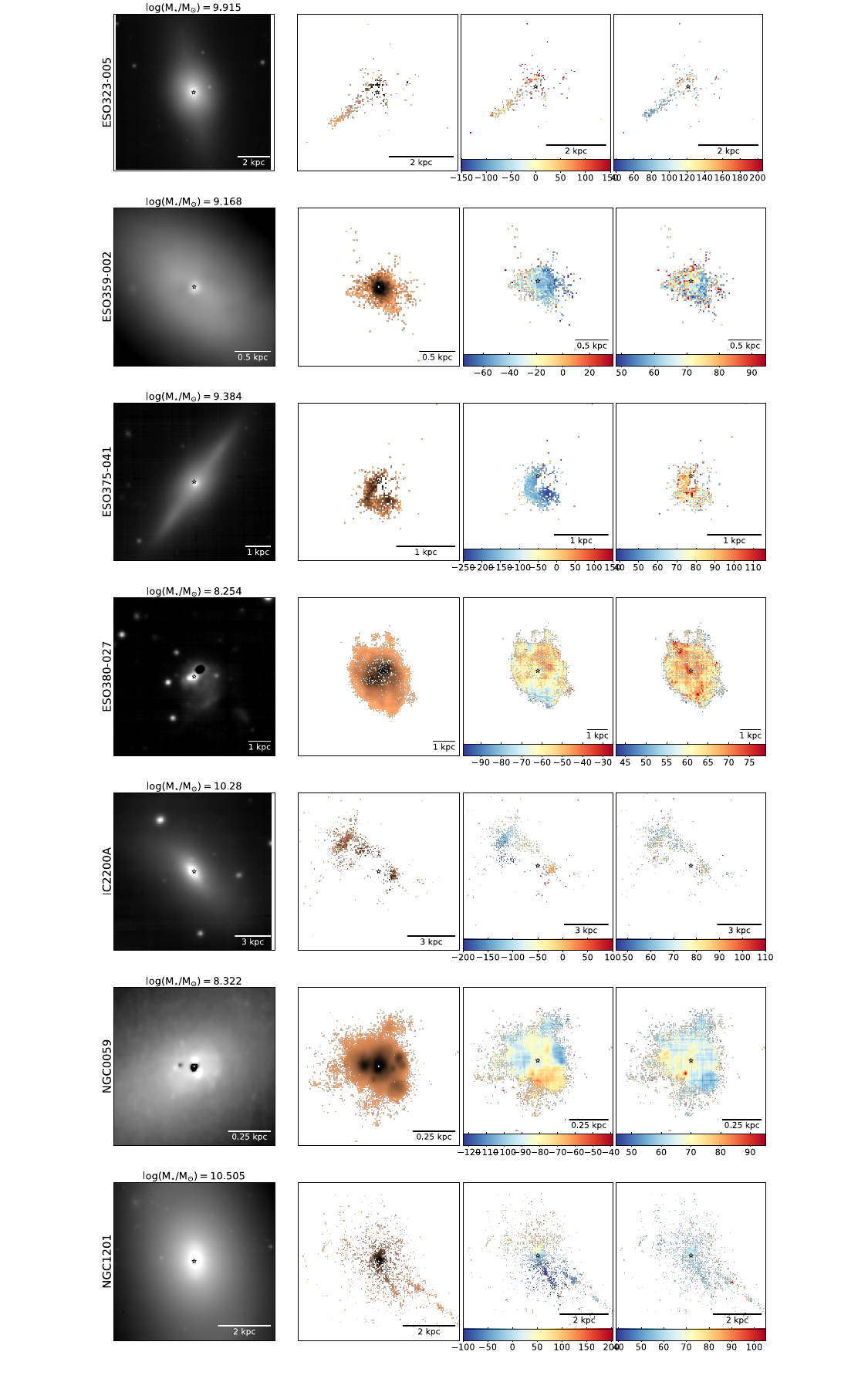}
  \caption{All 27 ETGs with filamentary H$\alpha$ morphologies. Each row of four panels corresponds to an individual source. The name (stellar mass) of the galaxy is given along the side (top) of the leftmost panel. The first, second, third, and fourth columns correspond to the optical continuum, H$\alpha$ flux, warm gas LOS velocity (km s$^{-1}$), and warm gas velocity dispersion (km s$^{-1}$), respectively. The physical scale is given in the bottom right of each panel.}
  \label{fig:fils}
\end{figure*}


\begin{figure*}
  \centering
  \includegraphics[width=0.8\linewidth]{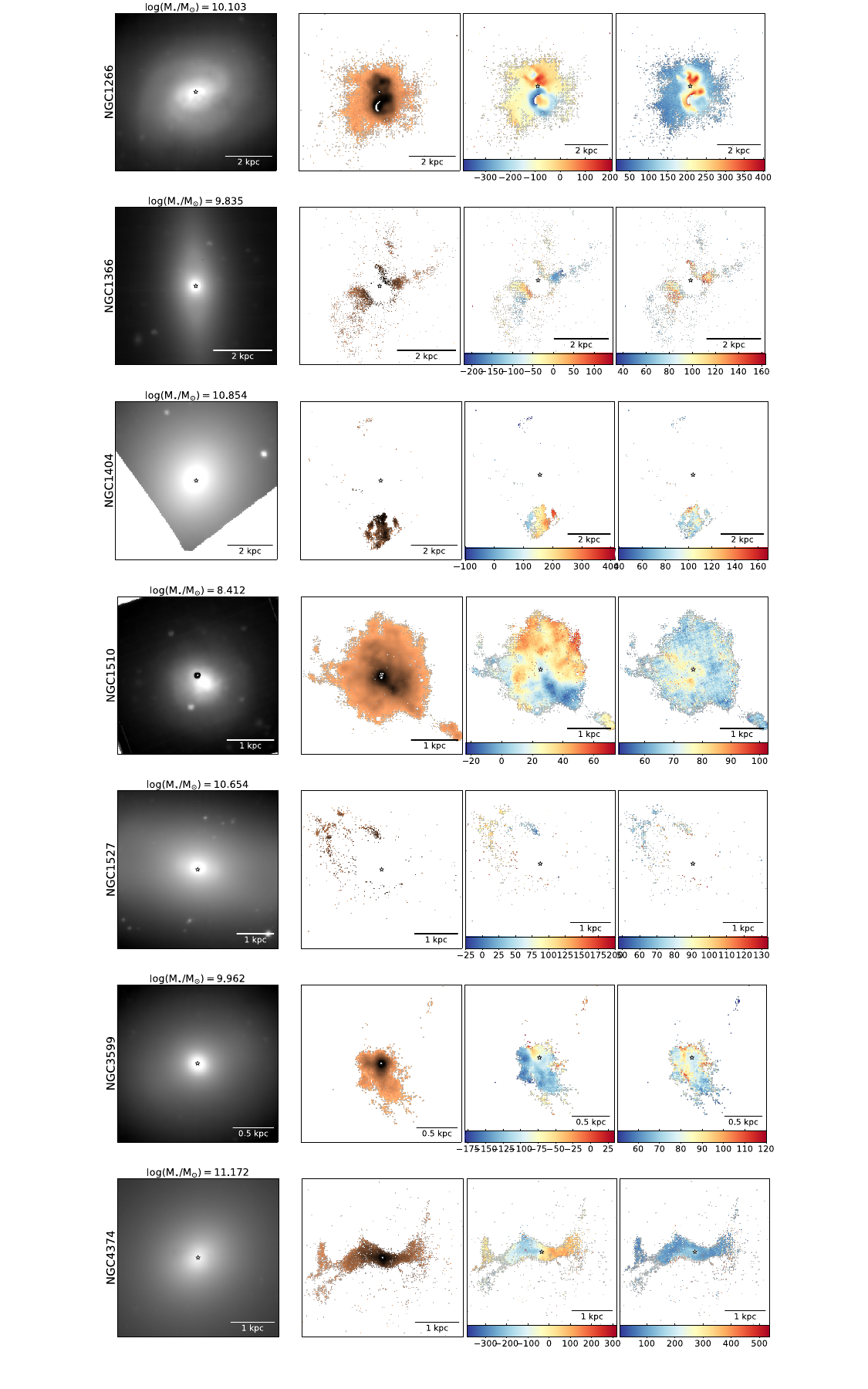}

  \medskip
  \par\small\emph{\figurename~\ref{fig:fils} (continued).}
\end{figure*}

\begin{figure*}
  \centering
  \includegraphics[width=0.8\linewidth]{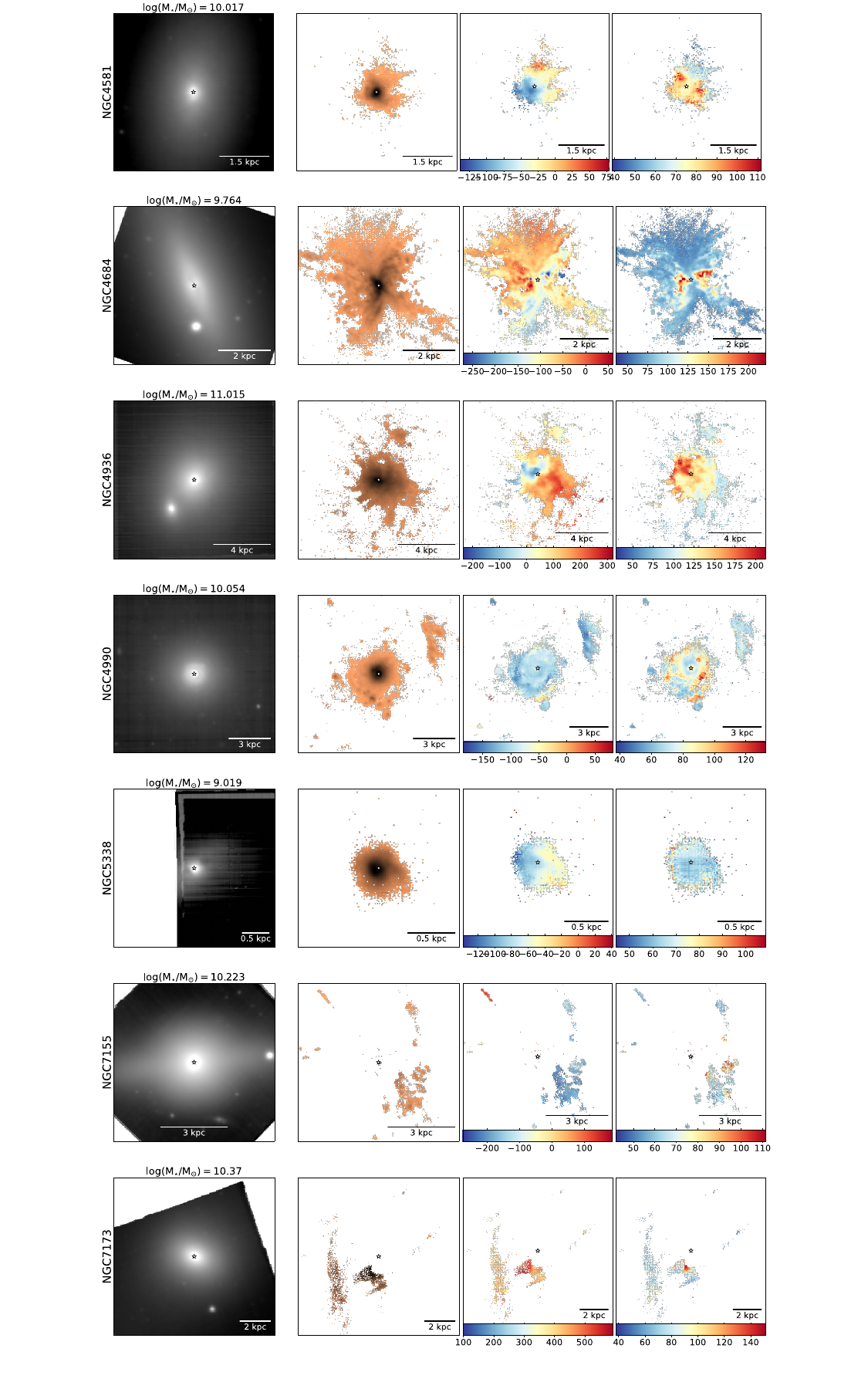}

  \medskip
  \par\small\emph{\figurename~\ref{fig:fils} (continued).}
\end{figure*}

\begin{figure*}
  \centering
  \includegraphics[width=0.8\linewidth]{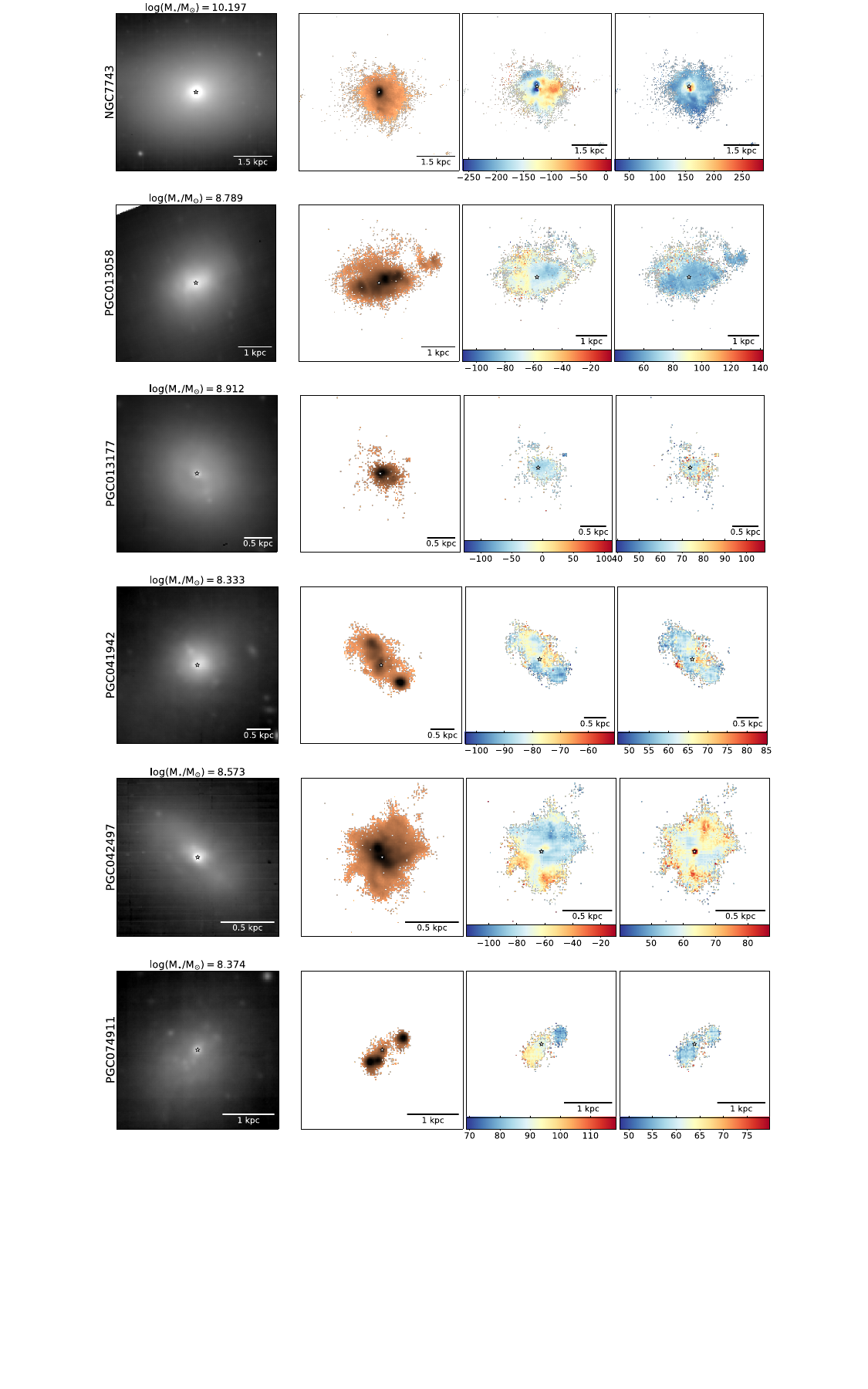}

  \medskip
  \par\small\emph{\figurename~\ref{fig:fils} (continued).}
\end{figure*}

\subsection{A case study with NGC~4374}
\label{subsec:N4374}

\begin{figure*}
    \centering
    \includegraphics[width=0.9\linewidth]{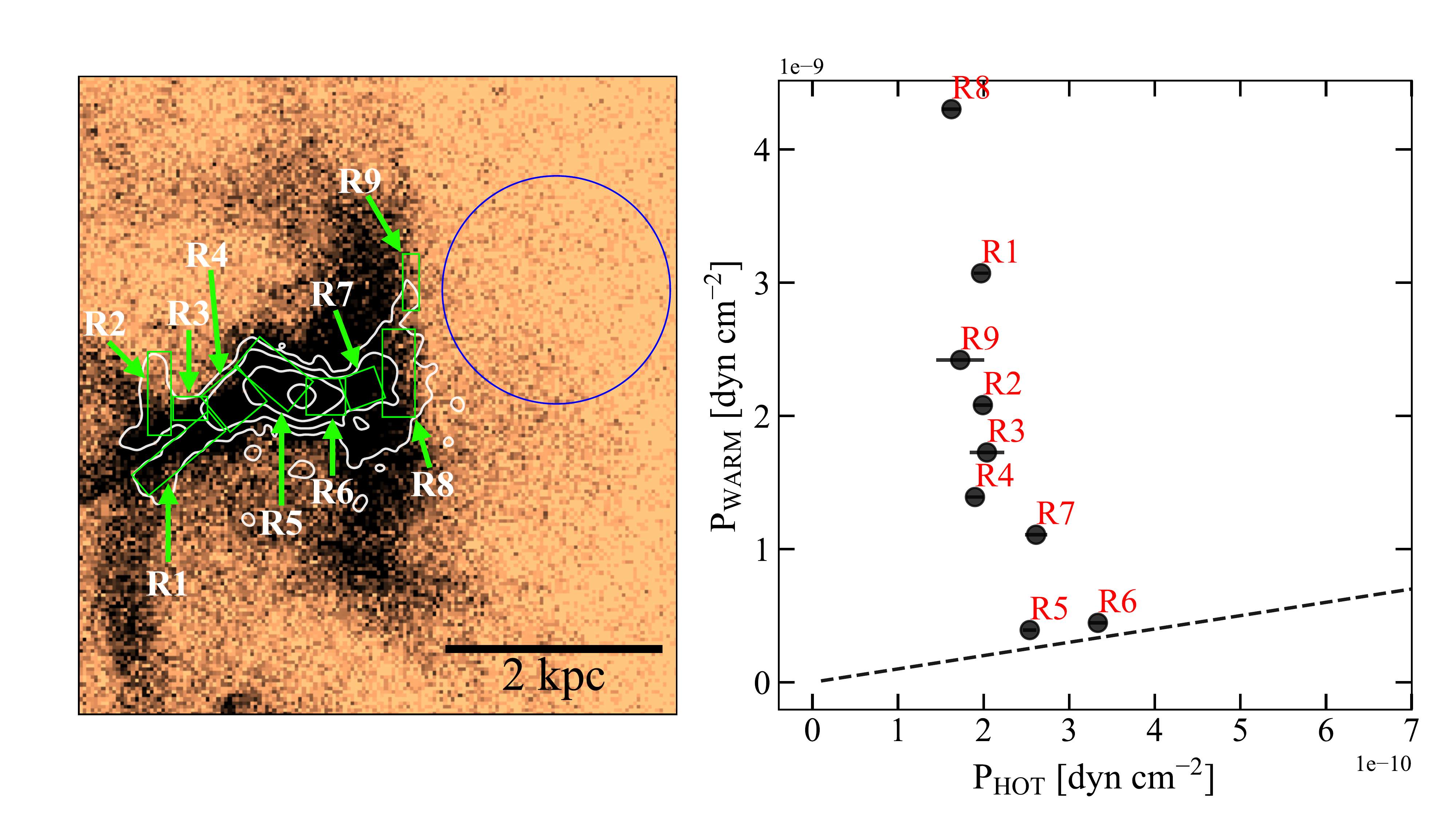}
    \caption{Left: \emph{Chandra} 0.5-2.0 keV image of NGC~4374 overlaid with the various regions used to extract source spectra (green boxes) and the background spectrum (blue circle). The warm ionized gas is traced by the white contours. Right: Electron pressure of the warm ionized gas (vertical axis) against the electron pressure of the hot gas (horizontal axis). The labels on each data point correspond to the regions overlaid on the left panel. The dashed black line is the one-to-one line. All regions contain warm ionized gas that has a higher thermal pressure than the hot phase. This likely suggests that the hot gas has additional non-thermal pressure support.}
    \label{fig:press}
\end{figure*}

The limited existing \emph{Chandra} observations and the low mass (X-ray faint) nature of many galaxies in our sample do not allow us to obtain their hot gas properties other than their X-ray luminosity. 
One exception is NGC~4374 (M84), one of the most massive galaxies in our sample. It is a non-central ETG in the Virgo Cluster containing substantial hot gas with X-ray cavities coincident with jet-inflated radio bubbles. 
Both H$\alpha$ filaments and X-ray filaments are prominent at its center.  
Warm ionized gas is found at the X-ray bright rim (X-ray filaments) of the cavities (see Fig.~\ref{fig:fils} and Fig.~\ref{fig:press}). NGC~4374 is considered an iconic cooling flow ETG and serves as a scaled-down version of cool-core BCGs. It has been observed extensively with \emph{Chandra} with a total exposure time of $\approx880$\,ksec. 
NGC~4374 allows us to perform an in-depth study with its multiphase filaments and probe the connection between warm and hot gas. 

We compared the H$\alpha$ and X-ray surface brightness in nine regions within NGC~4374's H$\alpha$ flux map (Fig.~\ref{fig:fils} and Fig.~\ref{fig:press}). For the X-ray surface brightness, we account for the background contribution using a local background as shown in Fig.~\ref{fig:press} (blue circle). We find $\mathrm{SB}_{X}/\mathrm{SB}_{{\rm H}\alpha}$ ratios ranging from $\sim1.48-22.3$. The X-ray and H$\alpha$ surface brightness values range from $7.05-149~(\times10^{38})$ and $0.840-68.6~(\times10^{38})$ erg s$^{-1}$ kpc$^{-2}$.


\begin{figure*}
    \centering
    \includegraphics[width=0.95\linewidth]{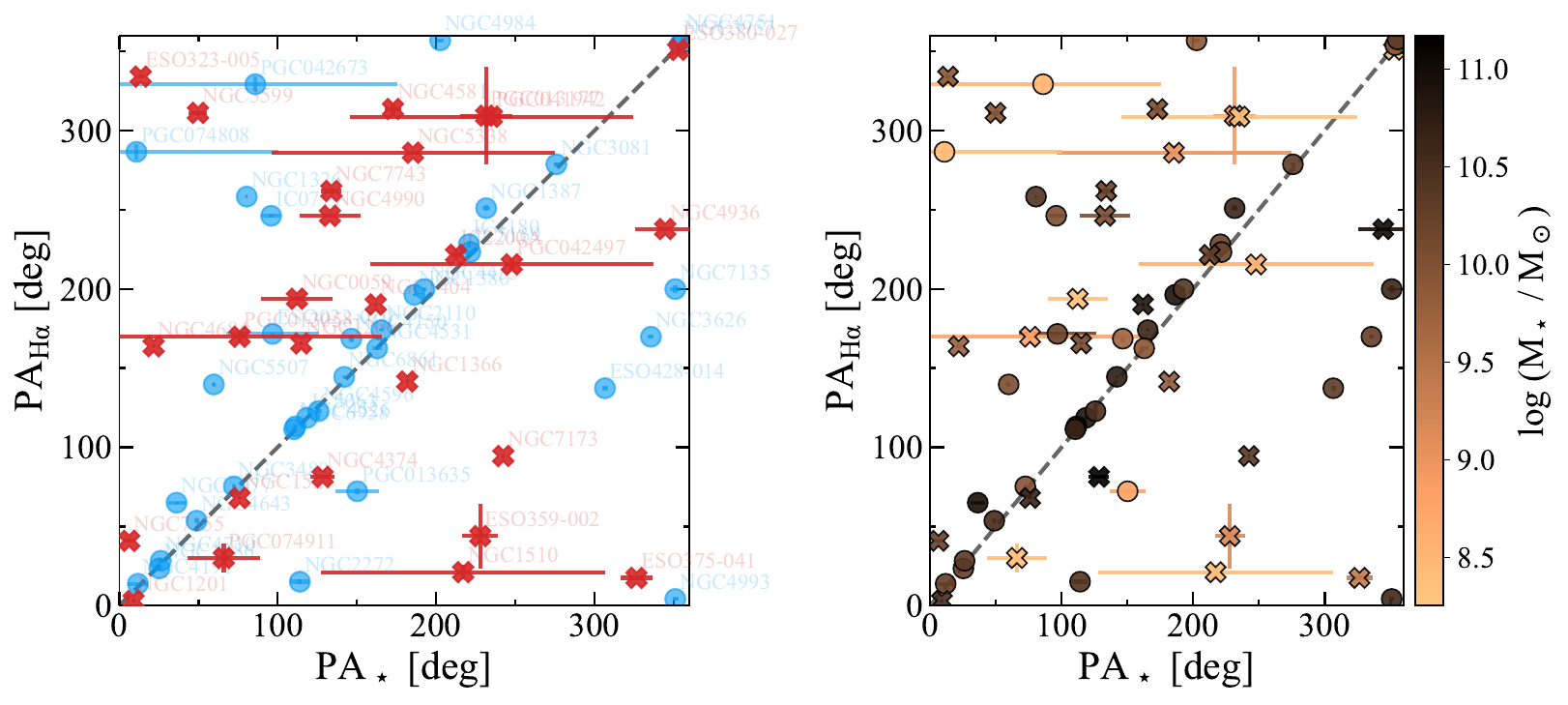}
    \caption{Left: The kinematic PA of the warm gas (vertical axis) vs. the kinematic PA of the stars (horizontal axis). Sources with filamentary H$\alpha$ nebulae are shown as red X's while rotating H$\alpha$ disks are blue circles. The dashed line represents the one-to-one relation. Right: Same axes as left plot, though color now corresponds to stellar mass as indicated by the right colorbar. X's represent H$\alpha$ filaments while circles represent rotating H$\alpha$ disks. The frequently observed misalignment observed in ETGs with warm filaments suggests that stellar mass loss is likely a subdominant process in forming their warm gas.}
    \label{fig:PA}
\end{figure*}

We further derived the electron pressure of each of the two phases within the same regions. To derive the pressure of the warm phase, we assume a uniform electron temperature $T_e=10^4$ K. The warm gas electron number density ($n_e$) can be computed at each spaxel using PyNeb \citep{luridiana2015pyneb}, given that spaxel's [SII]$\lambda$6717/[SII]$\lambda$6731 flux ratio. Note that variations in the warm gas temperature are expected, and should be considered when interpreting the derived $n_e$ under our isothermal assumption.

To determine the electron temperature and number density of the hot gas, we extracted the X-ray spectra from these regions (using the same local background described earlier). We then fit the spectra with an absorbed thermal plasma model, {\tt phabs}$\times${\tt apec}, in \verb|Xspec| version 12.12.1 \citep{arnaud1996xspec}. 
We derive $n_e$ of the hot gas using the fit's reported normalization parameter, $norm$, via:
\begin{equation}
\label{ne_hot}
    n_e = \sqrt{\frac{4\pi~10^{14}}{0.85 ~V}~D^2~(1+z)^2 ~norm}
\end{equation}
where $V = \pi a \left(b/2\right)^2$ is the volume of the filament, which is estimated by assuming that the gas enclosed by our 2D box regions (with side lengths a and b) used to extract the spectra, are approximately cylindrical. $D$ and $z$ are the angular diameter distance and redshift, respectively.

The electron pressure ($P_e$) for both phases is calculated using $P_e=n_ek_BT_e$. Fig.~\ref{fig:press} (right) compares the average $P_e$ between the warm and hot gas within each of the nine regions. The two inner-most regions (R5 and R6) contain warm and hot gas that are close to being in thermal pressure equilibrium. Though, on average, the electron pressure of their warm phase is a factor of $\sim1.4$ greater than that of the hot phase. Whereas for the extended regions (R1-R4, R7-R9), the electron pressure of the warm phase is significantly elevated compared to the hot phase by an average factor of $\sim12$. Thus, we find that the warm filaments have higher thermal pressure than that in the hot filaments. 
Our results are consistent when sampling various regions to extract the background contribution.

\bigskip
\section{Discussion} 
\label{sec:discuss}

The various mechanisms responsible for the formation of filamentary nebulae are poorly understood. Though, multiwavelength studies of BCGs suggest that condensation out of the hot phase is the primary origin. It is unclear if filaments in our sample of non-central ETGs follow the same evolutionary track of filaments in BCGs. In the following section we will interpret our spatially resolved MUSE and \emph{Chandra} data to put constraints on the origin of their filamentary warm gas. We will also discuss the feasibility of various physical processes driving their line emission.

\subsection{Formation channels of warm filaments}
\label{subsec:formation}

\subsubsection{Stellar mass loss}
\label{subsec:stellar}

Processes involving an ETG's stellar population can build up its warm ISM, specifically continual mass loss from post-asymptotic giant branch stars (pAGB) and planetary nebulae \citep{parriott2008mass,bregman2009mass,voit2011fate}. Material that is gradually shed from stars should be mostly shock-heated and mix in with the ambient, hot atmosphere. However, numerical hydrodynamics suggest that under the right conditions pertaining to the density of the hot ambient medium, as well as a star's velocity through that medium, can allow the shed material to dodge this shock heating thus preventing it from entering the hot phase \citep{parriott2008mass}. In this case, small, but non-negligible amounts of stellar mass loss will remain cool and be distributed within the ISM. 

If stellar mass loss is the dominant origin of warm gas, one would expect there to be spatial and kinematic overlap between the stars and gas. This can be observed by comparing the velocity fields of both populations. To do this, we calculate the kinematic position angle (PA) of both the stars and gas using the \verb|PaFit| Python package\footnote{https://pypi.org/project/pafit/} which implements the algorithm provided in the appendix of \cite{krajnovic2006kinemetry}. The PA allows for a consistent, geometric interpretation of the bulk motion of the stars and gas. Specifically, it is defined as the angle sweeping from the (projected) north to the location of the maximum positive velocity, in the counterclockwise direction. 

The left panel of Fig.~\ref{fig:PA} compares the kinematic PA of the stars ($\textrm{PA}_{\star}$) and gas ($\textrm{PA}_\textrm{gas}$). Table.~\ref{tab:results} provides the PA of both populations as well as their difference for each galaxy. Non-central ETGs with warm filaments are depicted as red x's and those with rotating warm disks are blue circles. The right panel has identical axes, as well as marker styles distinguishing between filaments and disks, though, it is now color-coded by stellar mass. The dashed black line in both panels shows the one-to-one line, which should be populated with ETGs that have warm gas primarily originating from stellar mass loss. 

Using the threshold of $| \textrm{PA}_{\mathrm{gas}} - \textrm{PA}_{\star} | > 30^{\circ}$ from \citet{lagos2015origin} to classify sources as having decoupled stars and nebulae, we find that 22/27 ($\sim81$$\%$) filamentary sources, show kinematic misalignment -- unlike the majority of the rotating disks. This is consistent with our analysis in \citet{eskenasy2024formation}, where we found that all non-central ETGs with warm filaments had widely offset PAs between the gas and stars, albeit with a significantly smaller sample size. 
The frequently observed misalignment observed in ETGs with warm filaments suggests that the two populations are generally decoupled from each other in this larger sample, hinting that stellar mass loss is likely a subdominant process in forming their warm gas.

\subsubsection{Cooling from the hot atmosphere}
\label{cooling}

The filamentary morphology and the decoupling between the warm gas and the stellar components resemble the warm ionized nebulae typically found in the BCGs of cool-core clusters, implying a similar formation mechanism. 
In BCGs, warm and cold filaments often appear to wrap around X-ray cavities, motivating the interpretation that low-entropy gas becomes thermally unstable as the cooling time of the hot gas becomes shorter than the free fall time 
as it is lifted in the updraft of rising X-ray bubbles -- promoting condensation to warm and cold filaments \citep{mcnamara2016mechanism}. 
The cold and warm gas can serve as fuel for the central supermassive black holes in BCGs. The black hole can in turn provide feedback to prevent the hot ICM from cooling at the predicted rate.

In a recent study of cool-core filaments, a close correlation has been identified between their X-ray surface brightness and H${\alpha}$ surface brightness -- SB$_{X}$ $=(3.33\pm0.34)$SB$_{\mathrm{H}\alpha}^{0.94\pm0.06}$ \citep{olivares2025halpha}. 
Strikingly, the slope of this relation matches exactly that found for the X-ray and H${\alpha}$ surface brightnesses of stripped tails of jellyfish galaxies \citep{sun2022universal}, demonstrating that the multiphase gas in both settings may share the same origin. The formation mechanism of the warm gas can occur in a broader astrophysical setting, and the presence of X-ray emitting hot gas is indispensable in this process.
As shown in Fig.~\ref{fig:Lx}, all filamentary ETGs are consistent with hosting a detectable hot-gas component, whereas not all rotating disk sources or systems without lacking H$\alpha$-emission do so. In particular, only one filamentary system lies near the expected contribution from unresolved LMXBs, but when accounting for both the measurement uncertainty and uncertainty in the LMXB relation, it is not significantly inconsistent with an excess above the unresolved LMXB-only expectation. This may imply that the presence of hot gas is a necessary but not sufficient condition for the formation of filamentary warm gas. We overlay NGC~4477 in Fig.~\ref{fig:Lx} (black square) as it is the smallest known system with X-ray cavities and cool-core signatures \citep{li2018x}. Although it was not observed with MUSE, it serves as a reference point demonstrating that the filamentary ETGs within our sample possess a higher hot gas content than NGC~4477\footnote{\cite{crocker2011molecular} found that the nucleus of NGC~4477 hosts a small-scale, rotating ring of molecular gas.}.

To further test the feasibility of cooling being the dominant formation channel of warm filaments, we estimated the timescales for producing their observed warm gas. We estimate the total mass of the warm gas using
\begin{equation}
\label{Mwarm}
    \mathrm{M}_{\mathrm{warm}} = 9.73 \times 10^8 (L_{\mathrm{H}\alpha,43})(n_{e,100}^{-1})~\mathrm{M_{\odot}}
\end{equation}
from Eq.~1 in \cite{nesvadba2007extreme}, assuming case B recombination \citep{osterbrock1989book}. $L_{\mathrm{H}\alpha,43}$ is the H$\alpha$ luminosity (in units of $10^{43}$ erg s$^{-1}$) and was extracted from the MUSE H$\alpha$ flux maps. $n_{e,100}$ is the electron density in units of 100 cm$^{-3}$ (determined using the procedure explained in Sec.~\ref{subsec:N4374}). 

To estimate the time for cooling to produce an ionized gas reservoir of this mass, we divide M$_\mathrm{warm}$ by the nominal cooling rate of the hot halo using
\begin{equation}
\label{Mcool}
    \dot {\mathrm{M}}_{\mathrm{cool}} =\frac{2}{5}\frac{\mu m_p L_X}{k_B T_X}
\end{equation}
adopted from \cite{su2017buoyant}. Here, $\mu=0.61$ is the average particle weight and $m_p$ is the proton mass. $L_X$ is the bolometric X-ray luminosity, which we estimated using our derived 0.5-2.0 keV luminosities provided in Table.~\ref{tab:results} and the energy conversion factor from \verb|Xspec| version 12.12.1. We assume $T_X\sim0.5$~keV for all sources, with the exception of NGC~1266 and NGC~4374 where we assume $kT=0.87$ keV and $kT=0.76$, respectively. 

\begin{figure}
    \centering
    \includegraphics[width=0.95\linewidth]{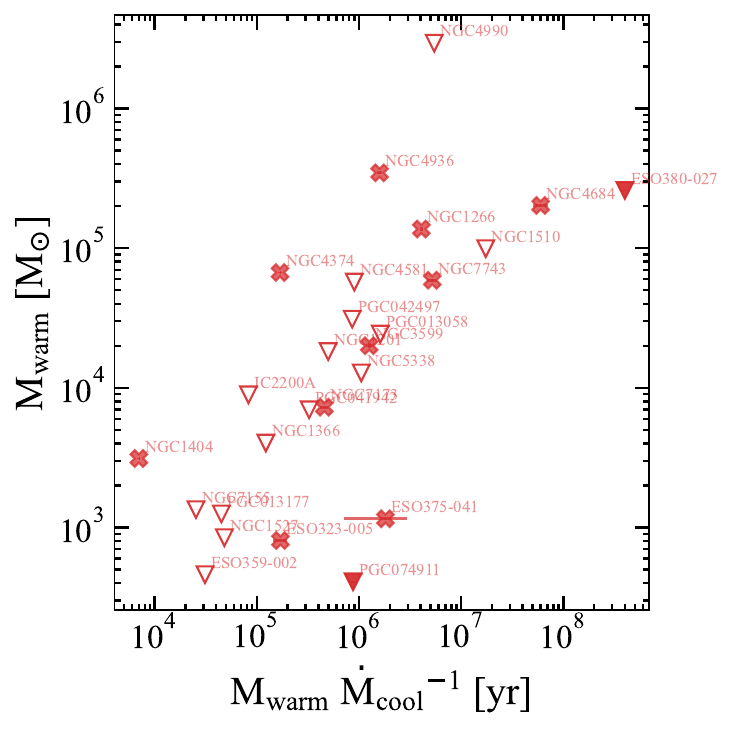}
    \caption{Total mass of the warm filaments versus the timescale to produce such nebulae through hot gas cooling. The crosses and inverted filled triangles show filamentary systems where $L_X$ was derived using \emph{Chandra} detections and upper limits, respectively. Inverted open triangles show systems where an upper-limit on $L_X$ was obtained from eRASS1, as they were not observed with \emph{Chandra}. We find that the timescales required to produce the filamentary nebulae through hot gas cooling are less than a Hubble time for all filamentary systems.}
    \label{fig:time}
\end{figure}

Fig.~{\ref{fig:time}} shows the ionized gas masses of all filamentary sources and the time required to produce such reservoirs via cooling. The crosses and inverted filled triangles show filamentary systems where $L_X$ was derived using \emph{Chandra} detections and upper limits, respectively. Inverted open triangles show filamentary systems not observed by \emph{Chandra}, and their upper-limit on $L_X$ is gathered from eRASS1 when available. Our derived timescales span $\sim10^{4-8}$ years, all less than the Hubble time. These estimates suggest that it is indeed possible for the filamentary warm gas reservoirs in our non-central ETGs to be produced through cooling.

To further investigate the cooling mechanism in non-central ETGs and its connection to BCGs and jellyfish galaxies, we performed an in depth study of NGC~4374, as it is the closest analogy of cool-core BCGs in our sample (see Sec.~\ref{subsec:N4374}). We compared the X-ray to H$\alpha$ surface brightness of the nine regions, finding a median $\mathrm{SB}_{X}/\mathrm{SB}_{{\rm H}\alpha}$ value of $\sim6.41$, which is larger than the BCG filaments and tails of jellyfish galaxies, differing by 9.06$\sigma$. As described in Sec.~\ref{subsec:N4374}, we derived the pressure of each gas phase and found on that, on average, the warm filaments, especially those extended to large radii, have higher electron pressure than that in the hot filaments. This result is opposite to what was found in \citet{olivares2025halpha} for multiphase filaments in cool-core BCGs. Given that the hot and warm gas should be roughly in pressure equilibrium with each other, \citet{olivares2025halpha} posited that the warm filaments have additional non-thermal pressure components. In the case of NGC~4374, additional pressure support is needed in the X-ray filaments. 

We investigate the role of non-thermal components such as turbulence and magnetic fields, that may be adding pressure support to NGC~4374's hot ISM. To estimate the turbulent pressure, we refer to \citet{ogorzalek2017improved} which computed the 1D turbulent velocity measurements of NGC~4374 and other massive ellipticals through line broadening and resonant scattering measurements. Interestingly, NGC~4374 was found to have one the lowest 1D velocity dispersions in their sample, with an upper limit of $v_{1\mathrm{D}}=47$ km~s$^{-1}$. The turbulent pressure ($P_{\mathrm{turb}}$) can be calculated with $P_{\mathrm{turb}}=\frac{1}{3}\rho v_{\mathrm{turb}}^2$ where $\rho$ is the gas mass density\footnote{$\rho=\mu n_em_p$ where $\mu$ is the mean molecular weight per electron. $n_e$ is the electron number density estimated in Eq.~\ref{ne_hot}. $m_p$ is the proton mass.} and $v_{\mathrm{turb}}=v_{1\mathrm{D}}\sqrt{3}$, assuming isotropic turbulence. This would bring the average hot gas turbulent pressure among the regions we consider to be $P_{\mathrm{turb}}\sim8.4\times10^{-12}$ dyn cm$^{-2}$. 

The magnetic pressure ($P_{\mathrm{B}}$) can be estimated as $P_{\mathrm{B}}=\frac{B^2}{8\pi}$ where $B$ is the magnetic field amplitude. We refer to \citet{finoguenov2001chandra} for the estimate of $B$. In their study, they estimated the strength of the magnetic field by determining the Faraday rotation from radio observations, as well as the electron number density of the hot gas which was extracted from the hot gas's three dimensional density profile using \emph{Chandra} X-ray observations. They find that the magnetic field associated with the X-ray plasma is particularly weak at $\sim0.8$ $\mu$G. This corresponds to $P_{\mathrm{B}}\sim2.5\times10^{-14}$ dyn cm$^{-2}$. Our estimates of the non-thermal pressure support added by turbulence and magnetic fields in the hot gas still does not account for its pressure imbalance with the warm gas. We posit that an additional non-thermal pressure component. For example, cosmic rays accelerated by NGC~4374's prominent AGN jet \citep{jenkins1977observations,finoguenov2001chandra} may be present. Although, the degree to which cosmic rays add pressure support is not well known. 

In the inner-most regions (R5 and R6), the average warm gas temperature required to be in thermal pressure equilibrium with the hot gas is $\sim7\times10^3$ K. In the extended regions (R1-R4, R7-R9), where the warm gas pressure is significantly elevated compared to the that of the hot gas, the average temperature of their warm gas would need to be $\sim1\times10^3$ K to be in pressure equilibrium with the hot phase. These temperatures are too low for a warm ionized medium, which is further support that the hot phase likely exhibits other pressure contributions.

We have demonstrated that while non-central ETGs are similar to BCGs in that their warm filaments originate from hot gas cooling, the physical processes within their ISM that likely regulate the evolution of multiphase gas may deviate from BCGs.

\subsubsection{Mergers and external accretion}
\label{external}
Kinematic misalignment between an ETG's warm gas and stars can be a key tracer of warm gas having an external origin such as galaxy mergers, tidal stripping from satellites, and accretion. \citet{davis2011atlas3d} found that it is common for local ETGs to exhibit this misalignment, thus demonstrating that a non-negligible fraction of warm gas in our sample of non-central ETGs is likely to be expected\footnote{Although subtle, but key differences in the merger histories -particularly timescales - arise from ranging environments, stellar masses, and whether an ETG is a fast or slow rotator.}. Our comparison of the kinematic position angles in Fig.~\ref{fig:PA} shows the majority of the filamentary systems have decoupled stars and gas suggesting an external origin of their warm gas. 

Although these non-central ETGs sit in environments much less dense compared to BCGs, it is plausible for them to have an extensive merger history. \citet{li2025origin} found that gas-rich quiescent galaxies are located in small groups or away from dense cluster cores. In these less dense regions, gas accretion is more efficient, perhaps due to these galaxies having lower velocity dispersions, leading to a higher probability of gas replenishment from gas-rich mergers \citep{mihos2004interactions}. Additionally, galaxies in dense cores may be too separated from the available intragalactic material to trigger cold mode accretion \citep{sharma2023h}. External accumulation of gas leading to the warm ionized filaments in our sample of non-central ETGs with misaligned gas and stars thus cannot be ruled out. 

\subsection{Powering of line emission}
\label{powering}

The traditional optical-line-diagnostic diagram proposed by Baldwin, Phillips and Terlevich (BPT; \citealp{baldwin1981classification}) is widely used to distinguish ionization sources in galaxies. However, it is known to have a variety of systematic uncertainties which can lead to ambiguous classification of these ionizing sources. Here, we adopt a modified optical-line-diagnostic diagram from \citet{ji2020constraining} to determine the ionization mechanisms in our sample. This diagram is a reprojected version of the BPT diagram, which ensures consistency of the model predictions among all three logarithmic line ratios: $\log([\mathrm{OIII}]\lambda5007/\mathrm{H}\beta)$ ($R$), $\log([\mathrm{NII}]\lambda6585/\mathrm{H}\alpha)$ ($N$), and $\log([\mathrm{SII}]\lambda\lambda6716,6731/\mathrm{H}\alpha)$ ($S$). The axes of this diagram are linear combinations of these three quantities, defined as

\begin{equation}
\label{P1}
    P_1 = 0.63 N + 0.51 S + 0.59 R
\end{equation}
and
\begin{equation}
\label{P2}
    P_2 = -0.63 N + 0.78 S
\end{equation}

\begin{figure*}
    \centering
    \includegraphics[width=0.85\linewidth]{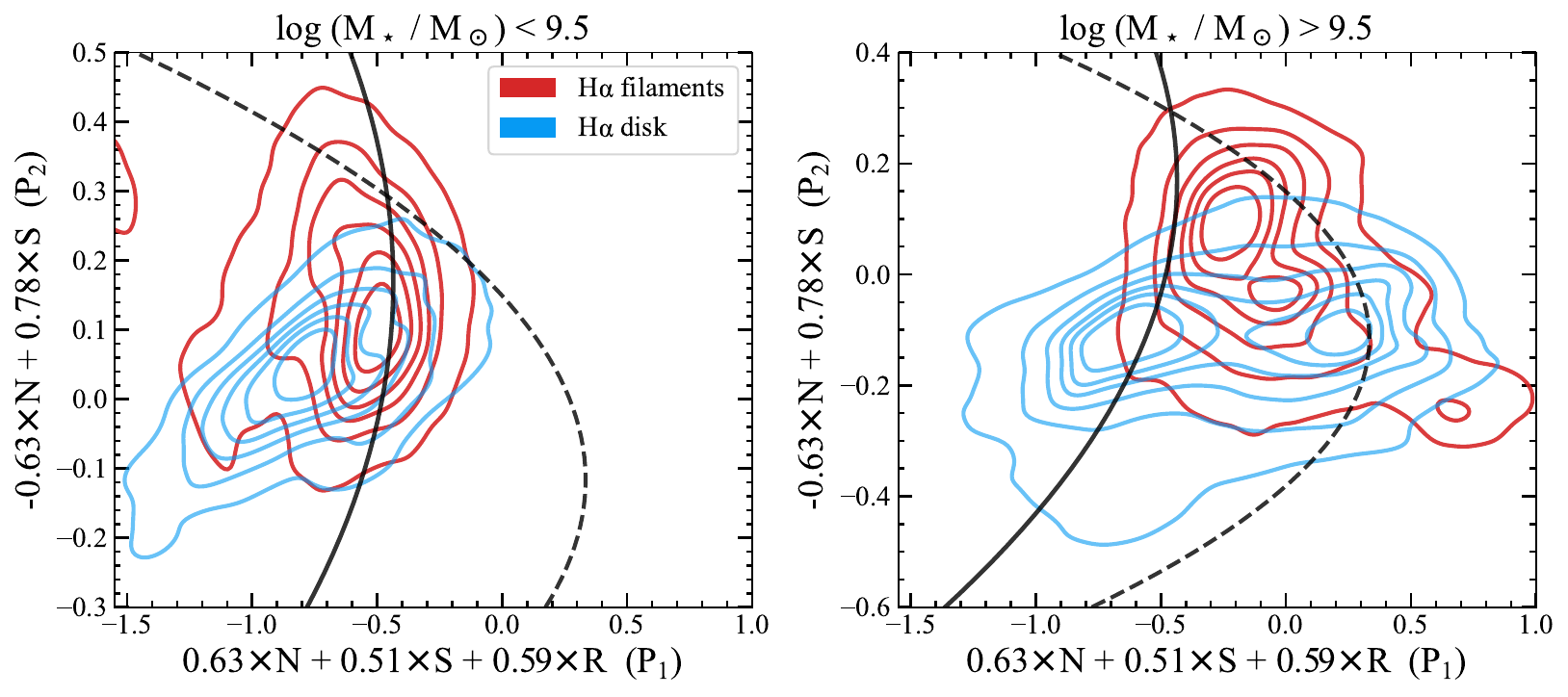}
    \caption{Top: Line-diagnostic diagram adopted from \citet{ji2020constraining}. Red (blue) contours depict the ETGs in our sample with H$\alpha$ filaments (rotating disks). Each set of contours are tracing individual spaxels within the emission line maps. The left (right) panel shows data only for ETGs with log(M$_{\star}$/M$_{\odot}$) $<$ ($>$) 9.5. The solid and dashed lines represent the demarcation lines presented in \citet{ji2020constraining}. Spaxels that fall to the left of the solid line suffer from $\lesssim$ 10\% AGN contribution, while spaxels to the right of the dashed line suffer from $\gtrsim$ 90\% AGN contribution - and thus we consider these to be demarcation lines separating SF, composite, and LINER/AGN spaxels.}
    \label{fig:reproj}
\end{figure*}

Fig.~\ref{fig:reproj} shows our systems placed in this reprojected diagram. As described in \citet{ji2020constraining}, spaxels that fall to the left suffer from less than 10 percent AGN contamination, while those to the right suffer from greater than 10 percent AGN contamination. Although the demarcation lines are created to represent this fractional AGN contribution, spaxels to the right may suffer from LINER contamination which can give rise to similar line ratios to that of Seyfert 2 emission. Thus, the rightmost part of the diagram is conservatively representing the broader family of LINERs, which may be powered by AGN. The region in between both demarcation lines is analogous to the \emph{composite} region in traditional line-diagnostic diagrams. 

The physical interpretation of line ratios lying within this composite region has long been debated. One could consider an individual cloud of gas that is ionized by comparable fractions of photons from young massive stars and AGN/LINERs. While theoretically possible, this would require very special conditions such as a fine-tuned geometry between the cloud and equally contributing ionization sources, such as an AGN and HII regions. Thus, a gas cloud likely has a single dominant ionization mechanism, and the placement of line ratios within the composite region is likely due to multiple gas clouds that cannot be spatially resolved. Under these assumptions, we assume that our composite region is made up of spatially unresolved AGN/LINER clouds as well as clouds ionized by young massive stars. 

In the following subsections we discuss various physical mechanisms responsible for the observed line ratios in the non-central ETGs with warm filaments.

\subsubsection{HII Regions}

Nebulae with emission line ratios falling to the left of the leftmost demarcation line are thought to be primarily photoionized by young, massive stars.
The peak of the distribution of spaxels corresponding to low mass ($\log(\mathrm{M}_{\star}/\mathrm{M}_{\odot}) < 9.5$) ETGs
 with filaments sit in this zone, slightly to the left of the boundary between the SF and composite zones. Whereas for the high mass ($\log(\mathrm{M}_{\star}/\mathrm{M}_{\odot}) > 9.5$) ETGs with filaments, their distribution peak is located in the composite region. 
 
These two distinct placements in the line-diagnostic diagram suggest that the photoionization of filamentary nebulae, due to young massive stars, is more significant in low mass ETGs. While this sheds light on the mechanism responsible for powering their emission line spectra, this perhaps also helps distinguish which systems have an internal (cooling hot gas) vs. external (gas rich accretion/merger) origin of their warm filaments. In the latter process, the accretion of fresh, cold gas would fuel star formation and provide radiation fields capable of ionizing surrounding gas. This process may be relevant in the low mass filamentary ETGs which sit in the leftmost region of Fig.~\ref{fig:reproj}. Consistent with this picture, ESO~375-041, the only ETG with warm gas filaments whose soft X-ray emission is not clearly above the unresolved-LMXB expectation, is also a low-mass ($\log(\mathrm{M}_{\star}/\mathrm{M}_{\odot}) = 9.384$) system. This suggests that at least some low-mass systems may host filaments without requiring a substantial hot-gas halo. Conversely, the warm gas filaments in the high-mass ETGs are more plausibly linked to a cooling origin, as this process is more viable in deeper gravitational potentials that are able to retain diffuse hot gas.

\subsubsection{AGN}
\label{agn}

In non star-forming galaxies, photoionization from AGN can produce line ratios similar to that of the diverse mechanisms powering LINERs. While AGN are certainly expected to photoionize gas restricted to the central regions (parsec scales), they cannot fully explain extended LINER emission. For example, \citet{yan2012nature} found an [OIII]/[SII] ratio that increased with radius. They determine that this is explained by the ionization parameter, or the ratio between ionizing flux and gas density, also increasing with radius. This is physically motivated by the gas density having a shallower dependence on increasing radius, than that of the inverse-square law that radiation follows. An outward increasing ionization parameter is nonphysical if AGN are assumed to be the dominant ionization mechanism. \citet{eracleous2010assessment} similarly deduce that AGN are inadequate at producing LINER-like spectra, though from an energetics standpoint.

If AGN photoionization is the dominant source of powering the filamentary warm nebulae in our non-central ETGs, then we would expect their distributions in $P_1-P_2$ to significantly shift if we remove spaxels residing in the nuclear region. To test this, we only checked spaxels located at a distance greater than 0.3$\times$R$_e$. In both the low and high mass filamentary sources, the peak of their extended spaxel distribution does not shift. Thus, AGN photons are likely not the main driver of the line emission in warm filaments.

\subsubsection{Gas shocks}
\label{shocks}

The role of gas shocks in powering the emission line spectra of ETGs cannot be ignored, given the wide range of phenomena that can produce them. Emission arising from collisional excitation within the heated downstream gas can result in LINER-like line ratios. If the shock is fast enough, the intense UV radiation field from the cooling downstream gas can photoionize the precursor gas leading to LINER/Seyfert-like spectra. Although, the relative importance of interstellar shocks in creating LINER-like spectra has long been debated. Shock models in \cite{yan2012nature} produced line ratios that had velocity dependence opposite of that suggested by observations. \cite{annibali2010nearby} found their shock models to reproduce just the central emission line ratios, which could be explained by nuclear processes such as jet-driven outflows. 

However, more recent studies using spatially resolved IFU data have challenged these previous findings. \cite{lee2024ionized}, using MaNGA observations of quiescent galaxies, measured temperatures using the auroral-to-strong line flux ratios and when comparing them to updated shock models, they found that there is strong evidence for interstellar shocks to be the ionization source of extended LINERs. To probe the presence of shocks in the warm ionized filaments of M87, the BCG of the Virgo cluster, \citet{boselli2019virgo} calculated shock-sensitive line ratios such as [OI]$\lambda6300$/H$\alpha$, [SII]$\lambda6716,6731$/H$\alpha$, and [NII]$\lambda6548,6583$/H$\alpha$. They found elevated values for all three of these line ratios close to the AGN and along parts of the filaments suggesting that these regions are mainly ionized by gas shocks. They further explored the relationship between the velocity dispersion of the gas and these line ratios, motivated by the expectation that shock-excited gas should display a strong positive relationship between them \citep{ho2014sami, rich2015galaxy}. This tight correlation was found most clearly using [OI]$\lambda6300$/H$\alpha$, consistent with findings in \citet{ho2014sami}. An extremely tight correlation is observed in the nuclear region close to AGN spanning the highest velocity dispersion values. Whereas the more extended gas shows a less steep, higher-scatter relation spanning the entire velocity dispersion range. 
\begin{figure*}
    \centering
    \includegraphics[width=0.9\linewidth]{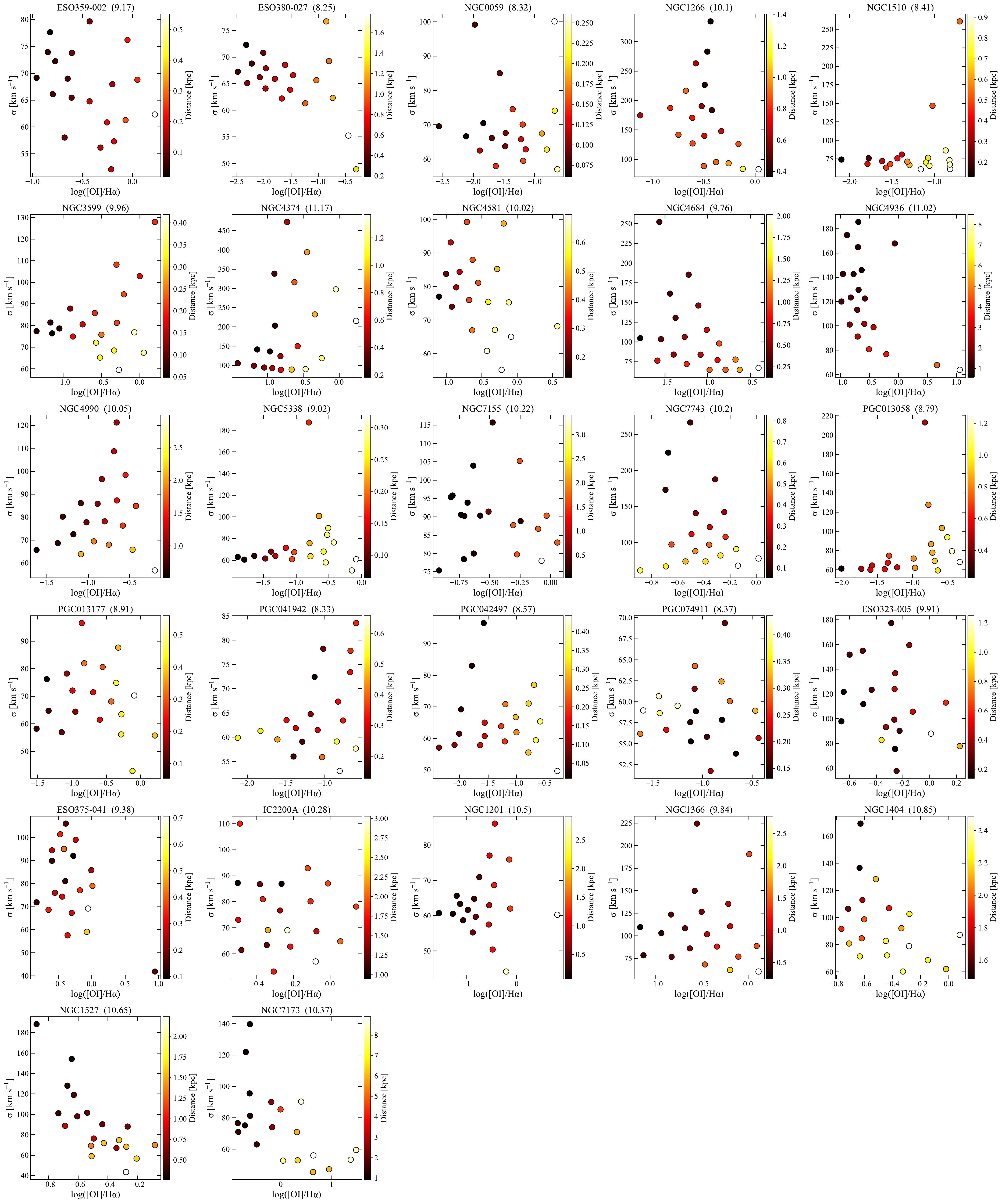}
    \caption{The logarithm of the [OI]$\lambda6300$/H$\alpha$ flux ratio (horizontal axis) vs. velocity dispersion (vertical axis) of the warm gas in all 27 non-central ETGs with warm filaments. Each data point is the median value of a K-means cluster. The color indicates the median radial distance of the spaxels with that cluster. The galaxy name is given in the top of each panel, and the logarithm of its stellar mass is in the parentheses. The extended gas on average does not exhibit a tight positive relationship between velocity dispersion and log([OI]$\lambda6300$/H$\alpha$), suggesting that gas shocks are likely not a dominant ionization source of the filamentary emission.}
    \label{fig:shock_diag}
\end{figure*}

We examine this relationship for all 27 of our filamentary ETGs in Fig.~\ref{fig:shock_diag}. To emphasize similar populations of spaxels, we apply K-means clustering to the data in each system. Each point in the figure represents the median of a K-means cluster in the $\sigma$-log([OI]$\lambda6300$/H$\alpha$) plane, and is color-coded by the median radial distance of the spaxels within that cluster. On average, these filamentary sources do not exhibit a tight correlation between velocity dispersion and [OI]$\lambda6300$/H$\alpha$, particularly in the more extended gas. While some systems show signatures of nuclear shocks (traced by centrally concentrated clusters with high velocity dispersion and line ratio values), it is likely not a dominant ionization mechanism across the bulk of the filamentary structures in our sample. Nevertheless, in the broader context of ETGs, we acknowledge that the role of gas shocks remains actively debated, as previous studies have demonstrated that shocks can reproduce LINER-like line ratios in extended gas \citep{lee2024ionized}.

\subsubsection{pAGB stars}
\label{pagb}

Photoionization from pAGB stars has long been thought to contribute to the emission line properties of ETGs \citep{trinchieri1991h,binette1994photoionization,stasinska2008can,lagos2022spatially}. These are viable candidates since their harder radiation fields compared to younger stars are capable of reproducing line ratios observed in composite regions and LINERs. A strong correlation between emission line flux and stellar luminosity within emission line regions in ETGs \citep{macchetto1996survey,sarzi2006sauron,sarzi2010sauron} may be compelling evidence for this. Such a correlation would indicate that the production site of ionizing photons is distributed similarly to the stellar population. However, more recent studies using MaNGA observations of ETGs have challenged the feasibility of pAGB stars in photoionizing extended LINER gas, given discrepancies found between model predictions and their observed line ratios and temperature estimates \citep{yan2018shocks,lee2024ionized}. 

For these reasons, we deduce that pAGB stars may have photoionized the warm filaments to some degree. However, given that the nebulae in these systems, unlike many ETGs with rotating gas disks, tend to be highly decoupled to the stellar population, an additional ionization source may be at play.

\subsubsection{Self-irradiated condensates}
\label{selfirrad}

An interesting, often overlooked, source of optical line emission may be directly related to multiphase gas condensing out of the hot ISM. As thermally unstable gas cools from $\sim10^{6-7}$~K down to $\sim10^4$~K, X-ray and EUV photons will be emitted. Additionally, as described back in Sec.~\ref{cooling}, lower temperature, higher density gas will condense out of the hot phase. The aforementioned high energy photons will irradiate the condensate, and can power its optical emission line spectrum. In this scenario, the hot gas and warm filaments should appear to be adjacent to each other. However, on average, our \emph{Chandra} exposures are not deep enough to resolve the hot gas morphology at the scales of the warm filaments. Indeed, for the two filamentary ETGs in this sample with sufficiently deep \emph{Chandra} observations, NGC~1266 and NGC~4374 (see Fig.~\ref{fig:press}), their hot gas appears to clearly overlap with their warm filaments (see \citealp{eskenasy2024formation} for more details on these two sources). Further evidence for spatial coincidence between the warm and hot gas phases in a subsample of non-cluster-central ETGs in the SAURON survey, including NGC~4374, is presented in \cite{sarzi2010sauron}. This observational connection is consistent with a broader, multiphase cooling framework that has been explored in detail particularly in cooling-flow clusters.  \citet{voit1990self} and \citet{donahue1991photoionization} have demonstrated how high energy photons from hot gas cooling could be reprocessed into H$\alpha$ radiation, and produce the observed warm gas emission line spectra (see also \citealp{ferland2009collisional,sarzi2010sauron,polles2021excitation}). 

This is an enticing solution to explain the seemingly unique ionization processes that power the warm filaments in our sample. Although these models are not well constrained, and effectively unexplored in non-central or isolated ETGs, it offers a physically motivated source of optical line emission in which AGN, shock, and pAGB star models cannot fully describe. In the most massive systems such as central cool-core group and cluster galaxies, there is substantial evidence pointing to a tight connection between cooling hot gas and warm filamentary nebulae \citep{olivares2019ubiquitous, polles2021excitation}. Naturally, this would reinvigorate hot gas cooling as a legitimate photoionizing source, now in the context of being directly connected to how filaments may form. As described in Sec.~\ref{cooling}, our analysis finds that cooling of the thermally unstable hot ISM is the primary process creating the observed warm filaments. We thus include hot gas cooling as a candidate mechanism for explaining the emission line spectra of our filamentary, non-central ETGs.

\section{Conclusions} 
\label{sec:conclusion}

Using spatially resolved MUSE IFU observations of 126 non-central ETGs, we find that 62 host warm ionized gas nebulae. Out of these 62 sources, 35 have gas in the form of rotating disks. The remaining 27 display filamentary emission, similar to what is observed in massive group and cluster central galaxies. These MUSE observations along with \emph{Chandra} X-ray observations allow us to infer that cooling from a thermally-unstable hot halo is a primary channel for forming the warm filaments observed in non-central ETGs, while external acquisition may also contribute in some systems. We also employ a set of thermodynamic and emission line analyses which provide insight into the physics within their ISM, thus allowing us to assess not only how filaments form, but what mechanisms regulate their existence and evolution. Our results can be summarized as:

\begin{enumerate}
  \item Our sample of 126 non-central ETGs, extracted from the 50MGC catalog, contains 62 that have warm ionized gas detected within their ISM. Of these, 35 host rotating warm gas disks and 27 host filamentary warm gas. A portion of the observed filaments are compactly situated around their optical center, while others are significantly extended. Their projected physical sizes range from $0.3-4$ kpc.
  \item Unlike the majority of non-central ETGs with rotating gas disks, most (22/27) filamentary sources have nebulae that are misaligned with their stellar population. Stellar mass loss thus is likely not a dominant formation channel of the filaments. This misalignment may suggest that external processes such as galaxy mergers and accretion have contributed to a fraction of their warm-ISM budget.
  \item All filamentary ETGs with archival \emph{Chandra} observations show soft X-ray emission that is consistent with exceeding (or, in one case, is indistinguishable from within the uncertainties) the expectation from unresolved LMXBs, indicating that systems with filamentary warm gas generally host an appreciable hot halo. We posit that thermal instabilities within this hot gas allow for cooler, denser phases, to condense out of it. This interpretation mirrors the leading model for filament formation in cool-core BCGs, and our results suggest that similar, cooling-driven pathways may be feasible in non-central ETGs that live in less extreme and less dense environments.
  \item NGC~4374, a non-central ETG within our sample has ample \emph{Chandra} coverage, as well as pronounced hot filaments, which allow us to measure their electron pressure. We compare the electron pressure between the warm and hot filaments and find that the warm filaments have a higher thermal pressure than that of the hot filaments. This is opposite as to what is found in BCGs, thus suggesting that while filaments similarly form between BCGs and non-central ETGs, the physical conditions of their ISMs may be significantly different. Thus, how the filaments are regulated over time may vary between the two systems. Additionally, we find a median X-ray$-$H$\alpha$ filament surface brightness ratio $~\sim9\sigma$ higher than that found in BCGs.
  \item Our emission line analysis using the optical line-diagnostic framework of \citet{ji2020constraining} indicates that different mechanisms are controlling the line emission in non-central ETGs with rotating gas disks and those with filaments. Standard, single-source explanations (such as AGN and gas shocks) appear to be unable to describe the line emission in the filaments. Photoionization from pAGB stars may be capable of powering the filaments, but a lack of spatial and kinematic overlap between the stars and extended filaments may suggest that this is not the sole mechanism. Given that we expect there to be significant hot gas in our filamentary sources, the high energy EUV photons and X-rays emanating from the cooling hot phase may be radiating their warm condensate, and, at least in part, power their emission lines. However, it is unclear if this is the dominant mechanism in all filamentary systems. This advocates for future studies to focus on the role of cooling and self-irradiation in explaining the long-standing puzzle of the observed ``composite'' and LINER-like emission in ETGs.
\end{enumerate}

\section*{Acknowledgments}

The authors thank the referee for their thorough and helpful comments on the manuscript. R.E., V.O., and Y.S. acknowledge support by NSF grant 2107711, NRAO grant SOSPA9-006, Chandra X-ray Observatory grants GO1-22126X, GO2-23120X, G01-22104X, NASA grants 80NSSC21K0714 and 80NSSC22K0856. V.O. acknowledges support from the DYCIT ESO-Chile Comite Mixto PS 1757, Fondecyt Regular 1251702, and DICYT PS 541.

\section*{Data Availability}
 
Data described within this article will be shared on request to the corresponding author.

\bibliographystyle{aasjournal}
\bibliography{MAIN}{}
\clearpage
\newpage
\mbox{~}
\clearpage
\newpage

\appendix
\setcounter{table}{0} 
\renewcommand{\thetable}{A\arabic{table}} 

\section{Multiwavelength observations and results}



\begin{longtable}{l c c c | c c c}
\caption{--  All MUSE IFU observations. (1) Galaxy name. (2) Right ascension in degrees. (3) Declination in degrees. (4) Distance in Mpc. Columns 2, 3, and 4 were gathered from the 50MGC catalog. (5) ESO Programme and Run Identification Codes. (6) Net exposure time of all object frames belonging to the Program ID in seconds. (7) Average FWHM seeing conditions over all object frames in arcseconds.}
\label{tab:muse}
\\

\hline
Galaxy & RA & Dec & Dist & Program ID & Exp. Time & Avg. Seeing \\
 & [$\degree$] & [$\degree$] & [Mpc] &  & [s] & [$^{\prime\prime}$] \smallskip\\
(1) & (2) & (3) & (4) & (5) & (6) & (7) \\
\hline
\endfirsthead
\multicolumn{7}{c}{\tablename\ \thetable\ -- \textit{Continued from previous page}}\\
\hline
Galaxy & RA & Dec & Dist & Program ID & Exp. Time & Avg. Seeing \\
 & [$\degree$] & [$\degree$] & [Mpc] &  & [s] & [$^{\prime\prime}$] \\
\hline
\endhead
\hline \multicolumn{7}{r}{\textit{Continued on next page}} \\
\endfoot
\hline
\endlastfoot

ESO~022-010 & 233.395354 & -78.123847 & 33.1 & 0103.D-0440(A) & 2405 & 1.90 \\
ESO~107-004 & 315.873333 & -67.181333 & 40.1 & 0101.D-0748(B) & 2805 & 2.85 \\
ESO~323-005 & 192.551084 & -41.514944 & 27.1 & 096.B-0325(A) & 2080 & 1.12 \\
ESO~351-030 & 15.038850 & -33.709000 & 0.1 & 0102.D-0372(A) & 2700 & 0.66 \\
 &  &  &  & 0101.D-0300(A) & 2700 & 0.68 \\
 &  &  &  & 105.208V.001 & 5400 & 0.53 \\
ESO~358-050 & 55.265084 & -33.779378 & 19.8 & 296.B-5054(A) & 5400 & 0.51 \\
ESO~358-059 & 56.264963 & -35.972658 & 19.7 & 296.B-5054(A) & 5400 & 0.72 \\
ESO~359-002 & 57.653136 & -35.909281 & 19.2 & 0104.A-0734(A) & 11400 & 0.80 \\
ESO~375-041 & 157.379187 & -35.259810 & 22.1 & 099.D-0022(A) & 2805 & 1.35 \\
ESO~380-027 & 186.444607 & -36.233767 & 33.1 & 0101.B-0703(A) & 2600 & 0.68 \\
ESO~428-014 & 109.130005 & -29.324806 & 23.2 & 097.D-0408(A) & 6160 & 0.84 \\
ESO~512-018 & 220.891379 & -24.460808 & 43.3 & 0103.A-0637(A) & 1320 & 2.75 \\
IC~0335 & 53.879325 & -34.447119 & 20.8 & 296.B-5054(A) & 3600 & 0.56 \\
IC~0719 & 175.077084 & 9.009869 & 31.2 & 095.B-0686(A) & 2520 & 1.03 \\
IC2200A & 112.026167 & -62.363111 & 45.2 & 106.2155.001 & 1320 & 0.89 \\
IC~3167 & 185.078294 & 9.545364 & 16.5 & 098.B-0619(A) & 1080 & 1.54 \\
 &  &  &  & 0100.B-0573(A) & 4048 & 1.05 \\
 &  &  &  & 106.218F.001 & 1860 & 0.96 \\
IC~3369 & 186.820533 & 16.024574 & 16.5 & 098.B-0619(A) & 3240 & 0.63 \\
IC~3637 & 190.081473 & 14.714951 & 16.5 & 0100.B-0573(A) & 7456 & 0.52 \\
IC~3735 & 191.335048 & 13.692744 & 17.1 & 0100.B-0573(A) & 3744 & 0.90 \\
IC~4180 & 196.735421 & -23.917055 & 37.5 & 0103.B-0582(A) & 12150 & 1.04 \\
IC~5063 & 313.009716 & -57.068830 & 44.1 & 60.A-9339(A) & 2400 & 0.84 \\
 &  &  &  & 60.A-9100(K) & 1200 & 0.84 \\
IC~5169 & 332.541705 & -36.088607 & 43.1 & 110.23ZH.001 & 4200 & 0.82 \\
IC~5250B & 341.842611 & -65.058574 & 43.6 & 0103.A-0637(B) & 1140 & 0.79 \\
NGC~0059 & 3.854700 & -21.444460 & 4.9 & 111.24UJ.002 & 3360 & 0.58 \\
NGC~0448 & 18.818841 & -1.626194 & 29.7 & 094.B-0225(A) & 10328 & 0.75 \\
NGC~1125 & 42.918500 & -16.650667 & 42.8 & 112.25YF.001 & 1800 & 0.57 \\
NGC~1201 & 46.033425 & -26.069691 & 20.7 & 108.227L.001 & 1760 & 0.65 \\
NGC~1266 & 49.003231 & -2.427347 & 28.5 & 0102.B-0617(A) & 2400 & 0.67 \\
NGC~1326 & 50.984918 & -36.464534 & 16.8 & 0100.B-0116(A) & 3600 & 1.26 \\
NGC~1351 & 52.645736 & -34.853968 & 19.2 & 296.B-5054(A) & 3600 & 0.71 \\
NGC~1366 & 53.473579 & -31.194110 & 18.6 & 0103.B-0331(A) & 8656 & 1.21 \\
NGC~1373 & 53.746657 & -35.171106 & 19.3 & 296.B-5054(A) & 5400 & 0.64 \\
NGC~1374 & 53.819099 & -35.226304 & 19.7 & 094.B-0298(A) & 590 & 1.46 \\
 &  &  &  & 296.B-5054(A) & 3600 & 0.71 \\
NGC~1375 & 53.820087 & -35.265648 & 19.8 & 296.B-5054(A) & 3600 & 0.50 \\
NGC~1379 & 54.015951 & -35.441232 & 18.6 & 296.B-5054(A) & 3600 & 0.36 \\
NGC~1380 & 54.114948 & -34.976093 & 21.0 & 296.B-5054(A) & 3600 & 0.65 \\
NGC~1380A & 54.197838 & -34.739581 & 19.9 & 296.B-5054(A) & 7200 & 1.04 \\
NGC~1381 & 54.132151 & -35.295159 & 21.8 & 099.B-0384(A) & 1224 & 2.16 \\
 &  &  &  & 296.B-5054(A) & 3600 & 0.70 \\
NGC~1386 & 54.192541 & -35.999249 & 15.9 & 094.B-0321(A) & 4000 & 0.82 \\
 &  &  &  & 296.B-5054(A) & 3600 & 1.02 \\
NGC~1387 & 54.237795 & -35.506630 & 19.1 & 296.B-5054(A) & 7200 & 1.04 \\
NGC~1389 & 54.298798 & -35.746046 & 21.2 & 296.B-5054(A) & 3600 & 0.62 \\
NGC~1396 & 54.527312 & -35.439947 & 19.9 & 094.B-0895(A) & 10400 & 0.87 \\
 &  &  &  & 0101.C-0329(C) & 6750 & 1.11 \\
NGC~1404 & 54.716312 & -35.594232 & 19.9 & 296.B-5054(A) & 3600 & 0.62 \\
NGC~1419 & 55.175514 & -37.510857 & 22.9 & 296.B-5054(A) & 5400 & 0.81 \\
NGC~1427 & 55.580861 & -35.392730 & 19.4 & 296.B-5054(A) & 4006 & 0.69 \\
NGC~1428 & 55.595467 & -35.153676 & 20.5 & 296.B-5054(A) & 5760 & 0.77 \\
NGC~1440 & 56.262084 & -18.266028 & 19.8 & 0104.B-0404(A) & 1760 & 0.76 \\
NGC~1460 & 56.557209 & -36.696360 & 19.8 & 296.B-5054(A) & 3600 & 0.62 \\
NGC~1510 & 60.885985 & -43.400089 & 9.5 & 110.243T.001 & 4808 & 0.74 \\
NGC~1527 & 62.100541 & -47.897028 & 18.2 & 096.D-0263(A) & 2220 & 0.97 \\
NGC~2110 & 88.047500 & -7.456181 & 29.3 & 108.2298.001 & 2520 & 1.96 \\
NGC~2272 & 100.672083 & -27.459472 & 30.0 & 0102.D-0095(A) & 2805 & 0.50 \\
NGC~2698 & 133.902206 & -3.183879 & 26.2 & 096.B-0449(A) & 8136 & 0.79 \\
NGC~2865 & 140.875759 & -23.161576 & 37.8 & 106.2155.001 & 150 & 0.43 \\
NGC~3081 & 149.873013 & -22.826332 & 32.3 & 099.B-0242(B) & 3600 & 0.64 \\
NGC~3115 & 151.308132 & -7.718498 & 9.7 & 60.A-9100(A) & 4861 & 0.79 \\
NGC~3412 & 162.722098 & 13.412161 & 11.3 & 0104.B-0404(A) & 1760 & 1.03 \\
NGC~3489 & 165.077357 & 13.901230 & 12.1 & 0104.B-0404(A) & 1760 & 0.65 \\
NGC~3585 & 168.321242 & -26.754877 & 20.4 & 094.B-0298(A) & 590 & 0.78 \\
NGC~3599 & 168.862291 & 18.110377 & 20.4 & 109.2332.001 & 2440 & 0.55 \\
NGC~3626 & 170.015830 & 18.356812 & 20.0 & 108.227L.001 & 1760 & 0.62 \\
 &  &  &  & 0104.B-0404(A) & 1760 & 1.33 \\
NGC~3957 & 178.506435 & -19.568753 & 20.3 & 110.24AS.004 & 5220 & 0.86 \\
NGC~4150 & 182.640252 & 30.401578 & 13.7 & 106.2174.003 & 3420 & 1.16 \\
NGC~4191 & 183.459961 & 7.200902 & 40.9 & 095.B-0686(A) & 2520 & 0.91 \\
NGC~4264 & 184.899027 & 5.846758 & 38.9 & 094.B-0241(A) & 6240 & 1.30 \\
NGC~4322 & 185.757182 & 15.905530 & 16.5 & 0100.B-0573(A) & 7922 & 0.98 \\
NGC~4365 & 186.116920 & 7.316747 & 23.3 & 094.B-0225(A) & 2502 & 1.40 \\
NGC~4371 & 186.230853 & 11.704233 & 17.0 & 60.A-9313(A) & 4200 & 0.82 \\
NGC~4374 & 186.265440 & 12.887109 & 18.5 & 0102.B-0048(A) & 2400 & 0.89 \\
NGC~4379 & 186.311393 & 15.607418 & 15.8 & 108.227L.001 & 1760 & 0.76 \\
NGC~4421 & 186.760618 & 15.461464 & 16.4 & 108.227L.001 & 1760 & 0.69 \\
NGC~4435 & 186.918655 & 13.078841 & 16.7 & 109.2332.001 & 2440 & 0.66 \\
NGC~4472 & 187.444536 & 7.999724 & 17.1 & 095.B-0295(A) & 600 & nan \\
NGC~4473 & 187.453607 & 13.429467 & 15.3 & 095.B-0686(A) & 2430 & 1.20 \\
NGC~4483 & 187.669357 & 9.015665 & 16.7 & 106.218F.001 & 5580 & 1.35 \\
NGC~4526 & 188.512545 & 7.699261 & 18.4 & 097.D-0408(A) & 2360 & 0.57 \\
NGC~4528 & 188.525353 & 11.321228 & 15.8 & 108.227L.001 & 1760 & 0.64 \\
NGC~4531 & 188.566191 & 13.075387 & 15.5 & 109.22VU.001 & 1760 & 1.21 \\
NGC~4578 & 189.377330 & 9.555088 & 16.3 & 109.22VU.002 & 1760 & 0.91 \\
NGC~4581 & 189.521536 & 1.477736 & 29.2 & 0103.A-0157(A) & 2000 & 0.48 \\
NGC~4596 & 189.983087 & 10.176163 & 16.9 & 109.22VU.001 & 1760 & 0.77 \\
 &  &  &  & 109.2332.001 & 2440 & 0.86 \\
NGC~4598 & 190.049768 & 8.383730 & 31.2 & 106.218F.001 & 8060 & 1.06 \\
NGC~4608 & 190.305291 & 10.155714 & 17.3 & 109.22VU.001 & 1760 & 0.81 \\
NGC~4612 & 190.386459 & 7.314876 & 16.6 & 109.22VU.001 & 1760 & 0.93 \\
NGC~4640 & 190.740746 & 12.286908 & 16.5 & 106.218F.001 & 1860 & 0.96 \\
NGC~4643 & 190.833783 & 1.978481 & 16.5 & 097.B-0640(A) & 3840 & 0.70 \\
NGC~4684 & 191.822867 & -2.727414 & 13.9 & 096.B-0449(A) & 9613 & 0.85 \\
NGC~4696D & 192.089880 & -41.714374 & 36.6 & 096.B-0325(A) & 1560 & 1.99 \\
NGC~4710 & 192.411541 & 15.164971 & 16.5 & 60.A-9307(A) & 2400 & 0.64 \\
NGC~4751 & 193.211683 & -42.659980 & 23.7 & 094.B-0298(A) & 295 & 0.85 \\
 &  &  &  & 095.B-0624(A) & 3600 & 1.73 \\
 &  &  &  & 105.20K2.001 & 10063 & 0.73 \\
NGC~4754 & 193.072886 & 11.313940 & 16.1 & 109.22VU.001 & 3520 & 1.05 \\
NGC~4936 & 196.070742 & -30.526288 & 38.0 & 094.A-0859(A) & 2460 & 1.00 \\
NGC~4984 & 197.238411 & -15.516305 & 12.9 & 097.B-0640(A) & 3921 & 0.60 \\
 &  &  &  & 097.D-0408(A) & 2500 & 1.97 \\
NGC~4990 & 197.322021 & -5.272778 & 44.8 & 106.21C7.002 & 2300 & 0.56 \\
NGC~4993 & 197.448751 & -23.383944 & 40.7 & 099.D-0668(B) & 2600 & 0.70 \\
NGC~5077 & 199.881825 & -12.656476 & 38.5 & 094.B-0298(A) & 295 & 1.55 \\
NGC~5206 & 203.433375 & -48.151153 & 3.2 & 60.A-9100(J) & 2400 & 0.68 \\
NGC~5338 & 208.360602 & 5.207755 & 12.8 & 097.D-0408(A) & 3000 & 0.92 \\
NGC~5507 & 213.332788 & -3.148870 & 25.6 & 096.B-0449(A) & 8136 & 1.10 \\
NGC~5770 & 223.312624 & 3.959756 & 19.2 & 096.B-0449(A) & 8136 & 1.20 \\
NGC~5846A & 226.621404 & 1.594962 & 33.6 & 097.A-0366(A) & 1760 & 0.76 \\
NGC~6861 & 301.831067 & -48.370281 & 28.6 & 095.B-0624(A) & 1800 & nan \\
 &  &  &  & 094.B-0298(A) & 295 & 0.84 \\
NGC~6958 & 312.177563 & -37.997322 & 33.1 & 60.A-9193(A) & 2040 & 1.66 \\
NGC~7135 & 327.441683 & -34.876315 & 35.7 & 0103.A-0637(A) & 1680 & 2.22 \\
NGC~7155 & 329.040494 & -49.521969 & 23.2 & 0103.B-0582(A) & 12140 & 0.79 \\
NGC~7173 & 330.513915 & -31.973746 & 31.3 & 099.A-0870(A) & 4830 & 1.36 \\
NGC~7743 & 356.088042 & 9.934028 & 20.3 & 109.2332.001 & 2440 & 0.68 \\
PGC~013058 & 52.784442 & -36.290115 & 19.2 & 296.B-5054(A) & 5400 & 0.72 \\
PGC~013177 & 53.391084 & -33.573224 & 20.1 & 296.B-5054(A) & 5400 & 0.62 \\
PGC~013343 & 54.226308 & -35.374671 & 19.4 & 296.B-5054(A) & 5400 & 0.82 \\
PGC~013449 & 54.805234 & -35.371474 & 18.9 & 096.B-0399(A) & 10400 & 1.03 \\
PGC~013635 & 55.689471 & -33.920214 & 15.0 & 0104.A-0734(A) & 3600 & 0.56 \\
PGC~039649 & 184.823289 & 5.875897 & 46.3 & 106.218F.001 & 7440 & 1.07 \\
PGC~041942 & 188.878577 & 6.333990 & 16.5 & 098.A-0364(A) & 2379 & 0.67 \\
PGC~042442 & 190.081986 & 15.935243 & 16.2 & 098.B-0619(A) & 1080 & 0.73 \\
 &  &  &  & 0100.B-0573(A) & 2024 & 1.13 \\
PGC~042497 & 190.209731 & 4.525812 & 16.5 & 0101.A-0168(A) & 2379 & 0.76 \\
PGC~042609 & 190.477541 & 9.584630 & 16.5 & 0100.B-0573(A) & 5592 & 0.90 \\
PGC~042673 & 190.588602 & 2.066666 & 16.5 & 099.A-0023(A) & 2379 & 3.38 \\
PGC~074808 & 54.580443 & -35.129094 & 17.6 & 094.B-0576(A) & 3000 & 2.20 \\
 &  &  &  & 096.B-0063(A) & 4800 & 1.44 \\
 &  &  &  & 097.B-0761(A) & 2400 & 1.05 \\
 &  &  &  & 098.B-0239(A) & 2400 & 1.36 \\
PGC~074811 & 54.589700 & -35.259694 & 35.1 & 096.B-0399(A) & 10400 & 0.81 \\
PGC~074879 & 55.141146 & -35.022956 & 18.5 & 0101.C-0329(C) & 2712 & 0.89 \\
PGC~074911 & 55.339041 & -33.769606 & 18.7 & 0104.A-0734(A) & 2400 & 0.77 \\
PGC~1073696 & 191.530242 & -3.269106 & 11.6 & 106.21A1.001 & 2628 & 1.03 \\
PGC~1193898 & 225.219120 & 1.404932 & 26.3 & 111.24KC.001 & 2628 & 1.69 \\
SDSSJ124649.44+024248.3 & 191.705892 & 2.713668 & 16.5 & 111.24KC.001 & 2628 & 1.48 \\
SDSSJ150033.02+021349.1 & 225.137583 & 2.230339 & 14.1 & 111.24KC.001 & 2628 & 1.71 \\
UGC07504 & 186.340058 & 16.429761 & 16.5 & 0100.B-0573(A) & 7922 & 0.74 \\
\end{longtable}

\clearpage

\clearpage

\begin{center}
\begin{longtable}{l c c}
\caption{--  All \emph{Chandra} X-ray observations. (1) Galaxy name. (2) Observation ID and instrument used during
observation. (3) Net exposure time integrated over all
observations.}
\label{tab:chandra}
\\
\hline
Galaxy & OBSID (Instrument) & Exp. Time \\
&&[ksec] \smallskip\\
(1) & (2) & (3) \\
\hline
\endfirsthead
\multicolumn{3}{c}{\tablename\ \thetable\ -- \textit{Continued from previous page}}\\
\hline
Galaxy & OBSID (Instrument) & Exp. Time \\
&&[ksec]\\
(1) & (2) & (3) \\
\hline
\endhead
\hline \multicolumn{3}{r}{\textit{Continued on next page}} \\
\endfoot
\hline
\endlastfoot

ESO~323-005 & 8179 (ACIS-S) & 29.8 \\
ESO~351-030 & 9555 (ACIS-S), 4716 (ACIS-S), 4717 (ACIS-S) & 175.5 \\
& 4713 (ACIS-S), 4718 (ACIS-S), 4714 (ACIS-S)& \\
& 4702 (ACIS-S), 4703 (ACIS-S), 4707 (ACIS-S)& \\
& 4715 (ACIS-S), 4712 (ACIS-S), 4704 (ACIS-S)& \\
& 4706 (ACIS-S), 4708 (ACIS-S), 4709 (ACIS-S)& \\
& 4710 (ACIS-S), 4711 (ACIS-S), 4698 (ACIS-S)& \\
& 4699 (ACIS-S), 4700 (ACIS-S), 4701 (ACIS-S)& \\
& 4705 (ACIS-S)& \\
ESO~358-050 & 18134 (ACIS-S) & 4.5 \\
ESO~358-059 & 18138 (ACIS-S) & 4.8 \\
ESO~375-041 & 23794 (ACIS-I), 24731 (ACIS-I) & 75.0 \\
ESO~380-027 & 27262 (ACIS-S), 25240 (ACIS-S) & 33.5 \\
ESO~428-014 & 4866 (ACIS-S), 18745 (ACIS-S), 17030 (ACIS-S) & 154.5 \\
IC~0335 & 18127 (ACIS-S) & 4.8 \\
IC~3735 & 8129 (ACIS-S) & 5.1 \\
IC~5063 & 21999 (ACIS-S), 22000 (ACIS-S), 22001 (ACIS-S) & 271.6 \\
& 22002 (ACIS-S), 21467 (ACIS-S), 21466 (ACIS-S)& \\
& 7878 (ACIS-S) & \\
IC~5169 & 10265 (ACIS-I) & 5.1 \\
NGC~1125 & 21418 (ACIS-S), 14037 (ACIS-I) & 58.1 \\
NGC~1266 & 11578 (ACIS-S),19498 (ACIS-S),19896 (ACIS-S) & 148.1 \\
&  11578 (ACIS-S),19498 (ACIS-S),19896 (ACIS-S) & \\
NGC~1351 & 18125 (ACIS-S) & 4.8 \\
NGC~1373 & 18139 (ACIS-S), 18123 (ACIS-S), 18133 (ACIS-S) & 14.4 \\
NGC~1374 & 18139 (ACIS-S), 18123 (ACIS-S), 18133 (ACIS-S) & 14.4 \\
NGC~1375 & 18139 (ACIS-S), 18123 (ACIS-S), 18133 (ACIS-S) & 14.4 \\
NGC~1380 & 9526 (ACIS-S) & 41.6 \\
NGC~1380A & 18128 (ACIS-S) & 4.9 \\
NGC~1381 & 4170 (ACIS-I) & 44.6 \\
NGC~1386 & 4076 (ACIS-S), 13185 (ACIS-S), 12289 (ACIS-S) & 100.4 \\
& 13257 (ACIS-S)& \\
NGC~1387 & 4168 (ACIS-I) & 45.6 \\
NGC~1389 & 4169 (ACIS-I) & 45.3 \\
NGC~1396 & 4172 (ACIS-I), 9530 (ACIS-S), 16639 (ACIS-S) & 364.0 \\
& 14527 (ACIS-S), 14529 (ACIS-S), 27748 (ACIS-S)& \\
& 26675 (ACIS-S), 49898 (ACIS-S), 240 (ACIS-S)& \\
& 2389 (ACIS-S), 319 (ACIS-S), 239 (ACIS-I)& \\
NGC~1404 & 9530 (ACIS-S), 16639 (ACIS-S), 14527 (ACIS-S) & 916.3 \\
& 14529 (ACIS-S), 4174 (ACIS-I), 9798 (ACIS-S)& \\
& 9799 (ACIS-S), 17549 (ACIS-S), 2942 (ACIS-S)& \\
& 16231 (ACIS-S), 17541 (ACIS-S), 17540 (ACIS-S)& \\
& 16233 (ACIS-S), 17548 (ACIS-S), 16232 (ACIS-S)& \\
& 16234 (ACIS-S), 27748 (ACIS-S), 26675 (ACIS-S)& \\
& 49898 (ACIS-S), 240 (ACIS-S), 2389 (ACIS-S)& \\
& 319 (ACIS-S), 239 (ACIS-I)& \\
NGC~1419 & 18132 (ACIS-S) & 4.8 \\
NGC~1427 & 4742 (ACIS-S) & 51.0 \\
NGC~1428 & 18135 (ACIS-S) & 4.8 \\
NGC~1460 & 18131 (ACIS-S) & 4.8 \\
NGC~2110 & 883 (ACIS-S) & 45.8 \\
NGC~2272 & 15668 (ACIS-S) & 29.7 \\
NGC~2865 & 2020 (ACIS-S) & 29.5 \\
NGC~3081 & 20622 (ACIS-S), 23657 (ACIS-S), 24378 (ACIS-S) & 261.8 \\
&  24379 (ACIS-S), 24380 (ACIS-S), 24381 (ACIS-S)& \\
& 24382 (ACIS-S), 24383 (ACIS-S), 26454 (ACIS-S)& \\
& 24384 (ACIS-S) & \\
NGC~3115 & 2040 (ACIS-S), 11268 (ACIS-S), 12095 (ACIS-S) & 1138.7 \\
& 13819 (ACIS-S), 13821 (ACIS-S), 13820 (ACIS-S)& \\
& 14383 (ACIS-S), 14419 (ACIS-S), 14384 (ACIS-S)& \\
& 13817 (ACIS-S), 13822 (ACIS-S)& \\
NGC~3412 & 4693 (ACIS-S) & 10.0 \\
NGC~3489 & 392 (ACIS-S) & 1.8 \\
NGC~3585 & 2078 (ACIS-S), 9506 (ACIS-S), 19332 (ACIS-S) & 215.5 \\
& 21034 (ACIS-S), 21035 (ACIS-S) & \\
NGC~3599 & 9556 (ACIS-S) & 19.9 \\
NGC~3957 & 9513 (ACIS-S) & 37.5 \\
NGC~4150 & 1638 (ACIS-S) & 1.7 \\
NGC~4322 & 23140 (ACIS-S), 23141 (ACIS-S), 6727 (ACIS-S) & 166.4 \\
& 12696 (ACIS-S), 14230 (ACIS-S), 9121 (ACIS-S) & \\
NGC~4365 &2015 (ACIS-S),5921 (ACIS-S),5922 (ACIS-S)&195.8 \\
&5923 (ACIS-S),5924 (ACIS-S),7224 (ACIS-S) & \\
NGC~4374 & 401 (ACIS-S),803 (ACIS-S),5908 (ACIS-S) & 884.1 \\
&6131 (ACIS-S),20539 (ACIS-S),20540 (ACIS-S) & \\
&20541 (ACIS-S),20542 (ACIS-S),20543 (ACIS-S) & \\
&21845 (ACIS-S),21852 (ACIS-S),21867 (ACIS-S) & \\
&22113 (ACIS-S),22126 (ACIS-S),22127 (ACIS-S) & \\
&22128 (ACIS-S),22142 (ACIS-S),22143 (ACIS-S) & \\
&22144 (ACIS-S),22153 (ACIS-S),22163 (ACIS-S) & \\
&22164 (ACIS-S),22166 (ACIS-S),22174 (ACIS-S) & \\
&22175 (ACIS-S),22176 (ACIS-S),22177 (ACIS-S) & \\
NGC~4379 & 8053 (ACIS-S) & 5.1 \\
NGC~4435 & 2883 (ACIS-S), 8042 (ACIS-S), 21376 (ACIS-S) & 125.0 \\
& 23037 (ACIS-S), 23189 (ACIS-S), 23200 (ACIS-S) & \\
NGC~4472 & 321 (ACIS-S), 322 (ACIS-I), 8107 (ACIS-S) & 577.5 \\
& 8095 (ACIS-S), 16261 (ACIS-S), 11274 (ACIS-S)& \\
& 16260 (ACIS-S), 16262 (ACIS-S), 12978 (ACIS-S)& \\
& 12889 (ACIS-S), 12888 (ACIS-S), 21647 (ACIS-S)& \\
 & 21648 (ACIS-S), 24981 (ACIS-S), 21649 (ACIS-S)& \\
NGC~4473 & 4688 (ACIS-S),11736 (ACIS-S),12209 (ACIS-S)& 107.6\\
NGC~4483 & 8061 (ACIS-S) & 5.1 \\
NGC~4526 & 3925 (ACIS-S) & 43.5 \\
NGC~4528 & 8054 (ACIS-S), 8112 (ACIS-S) & 10.2 \\
NGC~4531 & 2107 (ACIS-S) & 6.7 \\
NGC~4578 & 8048 (ACIS-S) & 5.1 \\
NGC~4596 & 11785 (ACIS-S), 2928 (ACIS-S) & 40.2 \\
NGC~4612 & 8051 (ACIS-S) & 5.1 \\
NGC~4640 & 11322 (ACIS-S) & 10.6 \\
NGC~4684 & 15190 (ACIS-I) & 28.8\\
NGC~4710 & 9512 (ACIS-S) & 29.8\\
NGC~4751 & 12957 (ACIS-S) & 7.5 \\
NGC~4754 & 8038 (ACIS-S) & 5.0 \\
NGC~4936 & 4997 (ACIS-I), 4998 (ACIS-I) & 28.9 \\
NGC~4993 & 19294 (ACIS-S), 18955 (ACIS-S), 24336 (ACIS-S) & 1567.5 \\
& 24887 (ACIS-S), 24888 (ACIS-S), 24889 (ACIS-S)& \\
& 23869 (ACIS-S), 23870 (ACIS-S), 24337 (ACIS-S)& \\
& 22677 (ACIS-S), 24923 (ACIS-S), 24924 (ACIS-S)& \\
& 26223 (ACIS-S), 20936 (ACIS-S), 20945 (ACIS-S)& \\
& 20860 (ACIS-S), 20938 (ACIS-S), 20937 (ACIS-S)& \\
& 21372 (ACIS-S), 21371 (ACIS-S), 21323 (ACIS-S)& \\
& 20728 (ACIS-S), 22158 (ACIS-S), 21322 (ACIS-S)& \\
& 22157 (ACIS-S), 22736 (ACIS-S), 22737 (ACIS-S)& \\
& 20861 (ACIS-S), 20939 (ACIS-S), 23183 (ACIS-S)& \\
& 23184 (ACIS-S), 23185 (ACIS-S), 21080 (ACIS-S)& \\
& 21090 (ACIS-S), 18988 (ACIS-S), 28523 (ACIS-S)& \\
& 28525 (ACIS-S), 28526 (ACIS-S), 28527 (ACIS-S)& \\
& 25734 (ACIS-S), 25527 (ACIS-S), 25528 (ACIS-S)& \\
& 27088 (ACIS-S), 29370 (ACIS-S), 29377 (ACIS-S)& \\
& 29395 (ACIS-S), 29397 (ACIS-S), 29398 (ACIS-S)& \\
& 29399 (ACIS-S), 27731 (ACIS-S), 27089 (ACIS-S)& \\
& 27090 (ACIS-S), 27091 (ACIS-S), 27752 (ACIS-S)& \\
& 27753 (ACIS-S), 27754 (ACIS-S), 28358 (ACIS-S)& \\
& 25733 (ACIS-S)& \\
NGC~5077 & 11780 (ACIS-S) & 28.7 \\
NGC~5206 & 28587 (ACIS-S), 28171 (ACIS-S) & 62.3 \\
NGC~5507 & 358 (ACIS-S),26398 (ACIS-S) & 18.7 \\
NGC~5846A & 4009 (ACIS-S), 7923 (ACIS-I), 788 (ACIS-S) & 149.9 \\
NGC~6861 & 3190 (ACIS-I), 11752 (ACIS-I) & 116.3 \\
NGC~7173 & 905 (ACIS-I) & 49.5 \\
NGC~7743 & 6790 (ACIS-S) & 13.8 \\
PGC~013343 & 4168 (ACIS-I), 4170 (ACIS-I) & 90.2 \\
PGC~013449 & 4176 (ACIS-I), 624 (ACIS-S) & 89.6 \\
PGC~039649 & 9569 (ACIS-S), 834 (ACIS-S) & 135.3 \\
PGC~042442 & 8084 (ACIS-S) & 5.2 \\
PGC~042609 & 2109 (ACIS-S) & 5.2 \\
PGC~074808 & 4173 (ACIS-I) & 45.1 \\
PGC~074811 & 4173 (ACIS-I) & 45.1 \\
PGC~074911 & 18134 (ACIS-S) & 4.5 \\
SDSSJ150033.02+021349.1 & 11324 (ACIS-S) & 6.3 \\
\end{longtable}
\end{center}

\clearpage

\clearpage

\begin{center}
\begin{longtable}{l c c c c c c c}
\caption{--  Optical and X-ray results for all sources with detected warm ionized gas. (1) Galaxy name. (2) Morphological classification of the warm gas. D and F correspond to rotating disks and filamentary nebulae, respectively. (3) Logarithm of the stellar mass from \citet{ohlson202350}. (4) Effective radius in kpc (arcseconds). (5) Kinematic PA of the warm gas nebulae in degrees. (6) Kinematic PA of the stellar population in degrees. (7) Absolute value of the difference between the gas and stellar kinematic PA. (8) \emph{Chandra} 0.5-2.0 keV luminosity within 1~R$_e$ of the optical center.}
\label{tab:results}
\\
\hline
Galaxy & Morph & log(M$_{\star}$/M$_{\odot}$) & R$_e$ & PA$_{\mathrm{gas}}$ & PA$_{\star}$ & $\Delta$PA & $L_X$ \\
&&&[kpc ($^{\prime\prime}$)]&[$\degree$]&[$\degree$]&[$\degree$]&[10$^{39}$ erg s$^{-1}$] \smallskip\\
(1) & (2) & (3) &(4) & (5) & (6) &(7) & (8) \\
\hline
\endfirsthead
\multicolumn{8}{c}{\tablename\ \thetable\ -- \textit{Continued from previous page}}\\
\hline
Galaxy & Morph & log(M$_{\star}$/M$_{\odot}$) & R$_e$ & PA$_{\mathrm{gas}}$ & PA$_{\star}$ & $\Delta$PA & $L_X$ \\
&&&[kpc ($^{\prime\prime}$)]&[$\degree$]&[$\degree$]&[$\degree$]&[10$^{39}$ erg s$^{-1}$]\\
(1) & (2) & (3) &(4) & (5) & (6) &(7) & (8) \\
\hline
\endhead
\hline \multicolumn{8}{r}{\textit{Continued on next page}} \\
\endfoot
\hline
\endlastfoot
ESO~022-010 & D & 10.195 & 1.47 (7.70) & 171.6 $\pm$ 0.5 & 96.8 $\pm$ 29.4 & 74.8 & - \\
ESO~323-005 & F & 9.915 & 1.03 (6.23) & 334.1 $\pm$ 0.9 & 13.6 $\pm$ 1.8 & 320.5 & 0.42 $\pm$ 0.06 \\
ESO~359-002 & F & 9.168 & 0.39 (3.86) & 43.9 $\pm$ 20.5 & 227.9 $\pm$ 11.3 & 184.0 & - \\
ESO~375-041 & F & 9.384 & 0.52 (4.27) & 17.5 $\pm$ 0.5 & 326.5 $\pm$ 9.9 & 309.0 & 0.06 $\pm$ 0.03 \\
ESO~380-027 & F & 8.254 & 0.12 (0.61) & 351.0 $\pm$ 2.4 & 352.8 $\pm$ 1.4 & 1.8 & $<$ 0.23 \\
ESO~428-014 & D & 10.295 & 1.68 (13.85) & 137.3 $\pm$ 0.6 & 306.6 $\pm$ 1.8 & 169.4 & 7.52 $\pm$ 0.11 \\
IC~0719 & D & 10.020 & 1.18 (9.05) & 246.2 $\pm$ 1.2 & 95.9 $\pm$ 5.9 & 150.3 & - \\
IC~2200A & F & 10.280 & 1.65 (7.34) & 221.5 $\pm$ 2.7 & 212.6 $\pm$ 3.2 & 8.9 & - \\
IC~4180 & D & 10.190 & 1.46 (7.09) & 228.2 $\pm$ 2.1 & 220.7 $\pm$ 0.9 & 7.5 & - \\
IC~5063 & D & 10.882 & 3.58 (15.00) & 118.6 $\pm$ 0.5 & 118.5 $\pm$ 0.9 & 0.1 & 27.23 $\pm$ 0.43 \\
IC~5169 & D & 10.180 & 1.45 (6.62) & 223.3 $\pm$ 0.6 & 221.6 $\pm$ 2.7 & 1.7 & 4.26 $\pm$ 1.15 \\
NGC~0059 & F & 8.322 & 0.13 (5.11) & 193.8 $\pm$ 1.5 & 112.2 $\pm$ 22.6 & 81.7 & - \\
NGC~1201 & F & 10.505 & 2.2 (18.88) & 3.0 $\pm$ 0.9 & 9.0 $\pm$ 0.9 & 6.0 & - \\
NGC~1266 & F & 10.103 & 1.31 (8.42) & 165.6 $\pm$ 0.6 & 114.9 $\pm$ 0.9 & 50.7 & 3.28 $\pm$ 0.1 \\
NGC~1326 & D & 10.407 & 1.94 (20.05) & 258.3 $\pm$ 0.5 & 80.5 $\pm$ 0.5 & 177.8 & - \\
NGC~1366 & F & 9.835 & 0.93 (10.21) & 141.5 $\pm$ 0.5 & 181.8 $\pm$ 0.9 & 40.3 & - \\
NGC~1380 & D & 10.907 & 3.69 (27.78) & 196.3 $\pm$ 0.6 & 186.3 $\pm$ 0.5 & 9.9 & 5.94 $\pm$ 0.16 \\
NGC~1386 & D & 10.248 & 1.58 (24.74) & 23.5 $\pm$ 0.6 & 25.3 $\pm$ 0.5 & 1.8 & 4.89 $\pm$ 0.07 \\
NGC~1387 & D & 10.614 & 2.53 (28.24) & 251.0 $\pm$ 1.8 & 231.6 $\pm$ 2.3 & 19.5 & 8.33 $\pm$ 0.19 \\
NGC~1404 & F & 10.854 & 3.45 (25.55) & 190.2 $\pm$ 0.5 & 161.9 $\pm$ 0.9 & 28.3 & 39.56 $\pm$ 0.09 \\
NGC~1510 & F & 8.412 & 0.15 (2.24) & 21.1 $\pm$ 4.5 & 217.1 $\pm$ 89.5 & 196.0 & - \\
NGC~1527 & F & 10.654 & 2.66 (32.59) & 68.0 $\pm$ 0.9 & 76.0 $\pm$ 0.5 & 8.0 & - \\
NGC~2110 & D & 10.631 & 2.59 (15.81) & 174.0 $\pm$ 0.6 & 165.5 $\pm$ 0.5 & 8.5 & 38.6 $\pm$ 0.46 \\
NGC~2272 & D & 10.413 & 1.95 (13.02) & 15.1 $\pm$ 1.8 & 114.0 $\pm$ 5.9 & 98.9 & 1.05 $\pm$ 0.12 \\
NGC~3081 & D & 10.316 & 1.72 (10.18) & 278.7 $\pm$ 0.9 & 275.9 $\pm$ 1.4 & 2.8 & 27.06 $\pm$ 0.42 \\
NGC~3489 & D & 9.960 & 1.09 (22.15) & 75.3 $\pm$ 0.9 & 72.4 $\pm$ 0.5 & 2.9 & 1.05 $\pm$ 0.16 \\
NGC~3599 & F & 9.962 & 1.09 (18.43) & 311.2 $\pm$ 1.8 & 49.7 $\pm$ 5.9 & 261.5 & 1.41 $\pm$ 0.11 \\
NGC~3626 & D & 10.253 & 1.59 (14.98) & 169.8 $\pm$ 0.5 & 335.6 $\pm$ 1.8 & 165.8 & - \\
NGC~3957 & D & 9.985 & 1.12 (9.70) & 354.0 $\pm$ 0.6 & 353.7 $\pm$ 0.5 & 0.3 & 0.15 $\pm$ 0.03 \\
NGC~4150 & D & 9.712 & 0.79 (49.02) & 168.6 $\pm$ 2.4 & 146.5 $\pm$ 1.4 & 22.0 & 0.88 $\pm$ 0.17 \\
NGC~4191 & D & 10.267 & 1.62 (8.68) & 13.8 $\pm$ 0.5 & 11.8 $\pm$ 6.3 & 2.10 & - \\
NGC~4374 & F & 11.172 & 5.2 (73.08) & 81.3 $\pm$ 0.5 & 128.4 $\pm$ 7.7 & 47.2 & 35.12 $\pm$ 0.1 \\
NGC~4435 & D & 10.368 & 1.84 (32.30) & 199.9 $\pm$ 1.2 & 192.7 $\pm$ 0.5 & 7.2 & 1.22 $\pm$ 0.04 \\
NGC~4526 & D & 11.030 & 4.33 (105.18) & 113.2 $\pm$ 0.6 & 111.3 $\pm$ 0.5 & 1.90 & 3.50 $\pm$ 0.09 \\
NGC~4531 & D & 9.927 & 1.04 (162.07) & 162.5 $\pm$ 1.2 & 162.8 $\pm$ 3.2 & 0.3 & 2.02 $\pm$ 0.15 \\
NGC~4581 & F & 10.017 & 1.17 (8.95) & 313.6 $\pm$ 3.0 & 172.8 $\pm$ 5.0 & 140.9 & - \\
NGC~4596 & D & 10.477 & 2.12 (15.99) & 122.8 $\pm$ 5.1 & 125.7 $\pm$ 2.3 & 2.9 & 0.75 $\pm$ 0.05 \\
NGC~4643 & D & 10.537 & 2.29 (24.44) & 53.6 $\pm$ 0.5 & 48.8 $\pm$ 1.4 & 4.70 & - \\
NGC~4684 & F & 9.764 & 0.85 (7.59) & 163.8 $\pm$ 0.5 & 21.7 $\pm$ 0.9 & 142.0 & 0.30 $\pm$ 0.05 \\
NGC~4710 & D & 10.336 & 1.77 (22.1) & 28.3 $\pm$ 1.5 & 26.2 $\pm$ 2.3 & 2.10 & 0.85 $\pm$ 0.06 \\
NGC~4751 & D & 10.283 & 1.65 (11.09) & 357.6 $\pm$ 1.5 & 355.5 $\pm$ 0.5 & 2.1 & 3.72 $\pm$ 0.36 \\
NGC~4936 & F & 11.015 & 4.24 (19.45) & 237.8 $\pm$ 0.5 & 344.6 $\pm$ 19.0 & 106.8 & 19.37 $\pm$ 0.9 \\
NGC~4984 & D & 9.874 & 0.97 (11.28) & 357.0 $\pm$ 0.5 & 202.6 $\pm$ 2.3 & 154.4 & - \\
NGC~4990 & F & 10.054 & 1.23 (5.50) & 246.2 $\pm$ 0.9 & 133.0 $\pm$ 19.4 & 113.3 & - \\
NGC~4993 & D & 10.565 & 2.38 (11.47) & 4.2 $\pm$ 0.9 & 351.0 $\pm$ 1.8 & 346.7 & 2.07 $\pm$ 0.06 \\
NGC~5077 & D & 10.947 & 3.89 (19.46) & 65.0 $\pm$ 0.6 & 36.2 $\pm$ 5.4 & 28.8 & 10.37 $\pm$ 0.48 \\
NGC~5338 & F & 9.019 & 0.32 (5.64) & 286.0 $\pm$ 1.5 & 185.4 $\pm$ 89.5 & 100.5 & - \\
NGC~5507 & D & 10.045 & 1.22 (9.29) & 139.7 $\pm$ 0.9 & 59.7 $\pm$ 0.9 & 80.0 & 0.51 $\pm$ 0.1 \\
NGC~6861 & D & 10.750 & 3.02 (15.12) & 144.5 $\pm$ 0.5 & 142.0 $\pm$ 0.5 & 2.5 & 21.2 $\pm$ 0.35 \\
NGC~6958 & D & 10.651 & 2.65 (13.81) & 111.4 $\pm$ 1.2 & 110.4 $\pm$ 0.9 & 1.0 & - \\
NGC~7135 & D & 10.622 & 2.56 (13.31) & 199.9 $\pm$ 2.4 & 351.0 $\pm$ 3.2 & 151.1 & - \\
NGC~7155 & F & 10.223 & 1.53 (10.95) & 40.9 $\pm$ 0.5 & 6.3 $\pm$ 3.2 & 34.6 & - \\
NGC~7173 & F & 10.37 & 1.85 (10.69) & 94.5 $\pm$ 0.5 & 242.4 $\pm$ 0.9 & 147.9 & 1.41 $\pm$ 0.13 \\
NGC~7743 & F & 10.197 & 1.48 (12.18) & 261.9 $\pm$ 2.1 & 133.9 $\pm$ 6.3 & 128.0 & 1.0 $\pm$ 0.11 \\
PGC~013058 & F & 8.789 & 0.24 (1.87) & 169.8 $\pm$ 3.0 & 76.0 $\pm$ 89.5 & 93.8 & - \\
PGC~013177 & F & 8.912 & 0.28 (2.89) & 309.4 $\pm$ 30.7 & 231.6 $\pm$ 16.3 & 77.9 & - \\
PGC~013635 & D & 8.807 & 0.25 (2.77) & 72.2 $\pm$ 2.1 & 150.2 $\pm$ 13.6 & 77.9 & - \\
PGC~041942 & F & 8.333 & 0.13 (1.17) & 308.8 $\pm$ 0.9 & 235.2 $\pm$ 89.5 & 73.7 & - \\
PGC~042497 & F & 8.573 & 0.18 (3.56) & 215.5 $\pm$ 3.3 & 247.8 $\pm$ 89.5 & 32.3 & - \\
PGC~042673 & D & 8.494 & 0.16 (1.76) & 329.3 $\pm$ 5.1 & 85.9 $\pm$ 89.5 & 243.4 & - \\
PGC~074808 & D & 8.422 & 0.15 (1.48) & 286.6 $\pm$ 4.8 & 10.9 $\pm$ 89.5 & 275.7 & $<$ 0.005 \\
PGC~074911 & F & 8.374 & 0.14 (1.33) & 30.1 $\pm$ 9.0 & 66.0 $\pm$ 23.1 & 35.9 & $<$ 0.13\\
\end{longtable}
\end{center}

\end{document}